\documentclass[twocolumn,prx,epsfig,rotate,noshowpacs,cjk,superscriptaddress,english]{revtex4}
\usepackage{graphicx}
\usepackage{amsmath}
\usepackage{dcolumn}
\usepackage{epsfig}
\usepackage{bm}
\usepackage{mathrsfs}
\begin{document}

\title{Hilbert Transform: Mapping Classical to Quantum Dynamics}

\author{Daxing Xiong}
\email{phyxiongdx@fzu.edu.cn}
\affiliation{Department of Physics,
Fuzhou University, Fuzhou 350108, Fujian, China}
\affiliation{Department of Physics,
Institute of Nanotechnology and Advanced Materials, Bar-Ilan
University, Ramat-Gan, 52900, Israel}

\author{Felix Thiel}
\email{thiel@posteo.de}
\affiliation{Department of Physics,
Institute of Nanotechnology and Advanced Materials, Bar-Ilan
University, Ramat-Gan, 52900, Israel}

\author{Eli Barkai}
\email{Eli.Barkai@biu.ac.il}
\affiliation{Department of Physics,
Institute of Nanotechnology and Advanced Materials, Bar-Ilan
University, Ramat-Gan, 52900, Israel}

\begin{abstract}
We propose a simulation strategy which uses a classical device of
linearly coupled chain of springs to simulate quantum dynamics, in
particular quantum walks. Through this strategy, we obtain the
quantum wave function from the classical evolution. Specially, this goal
is achieved with the classical momenta of the particles on the chain
and their Hilbert transform, from which we construct the many-body
momentum and Hilbert transformed momentum pair correlation functions
yielding the real and imaginary parts of the wave function,
respectively. With such wave function, we show that the classical
chain's energy and heat spreading densities can be related to the
wave function's modulus square. This relation indicates a
concept of ``phonon random walks'', and thus it provides a new
perspective to understand ballistic heat transport. The results here may give a
definite answer to Feynman's idea of using a classical device to
simulate quantum physics.
\end{abstract} \maketitle

\section{Introduction}
In his pioneering work entitled ``Simulating Physics with
Computers''~\cite{Feynman}, R. Feynman posed two important
questions. First: What kind of computers are we going to use to
simulate physics? This led to the fundamental concept of a quantum
computer~\cite{QC}. The second was: Can a quantum system be
simulated by a classical computer? The answer to this in the words
of Feynman is: ``No! Since this is called the hidden-variable
problem: it is impossible to represent the results of quantum
mechanics with a classical universal device''. Feynman also
conditions this statement (see precise details in~\cite{Feynman})
and writes that such rather far reaching conclusion is valid
provided that there is no ``hocus-pocus''. The aim of the present
work is to theoretically build a classical device that can be used
to simulate quantum dynamics. Our device is a system of springs
initially prepared at thermal equilibrium. The main difficulty
recognized by Feynman is, as he wrote: ``... we cannot simulate
$\psi$ in the normal way'', where $\psi$ is the wave function.
Indeed, quantum mechanics is built on interference and a complex
valued field. Of course, the interference is a wave property and for
that reason we can imagine that vibrations are useful. But how can
we get the complex valued wave function? Well, as we show below, the
hocus-pocus is based on the Hilbert transform~\cite{Hilbert}. The
Hilbert transform takes a real function $f(t)$ and extends it to the
complex plane in such a way that it satisfies the Cauchy-Riemann
equations~\cite{CR}. It has been used extensively extending real
signals to the complex plane. Our goal below is to identify what
kind of classical object which together with its Hilbert transform
gives $\psi$. In that sense the main focus of the paper is Feynman's
second problem on how to go from a classical device to quantum
mechanics. However, our work is also potentially related to his
first vision. Namely, the field of phononics~\cite{phononics}
suggests to use phonons as a source of computation, thus our work
shows precisely how this strategy can be used to simulate quantum
physics~\cite{QW-1,QW-2,QW-3,QW-4,QW-5,QW-6}.

In the early days of quantum theory, there have been several
relevant attempts to interpret quantum reality within a classical
framework~\cite{Bohm-1,Bohm-2,Nelson,Vink,Discussion1979,Discussion1994}.
However, this is not the goal of the present paper which is focused on
building a device. These attempts, Bohm's causal
interpretation~\cite{Bohm-1,Bohm-2} and Nelson's stochastic
approach~\cite{Nelson}, yield a recipe for quantum dynamics based on
classical concepts. These while sharing some general themes with our
construction, as we will show below, are never the less very
different from what we propose (see discussions in the conclusion
part). In fact, from the start, our goal is not to claim that
quantum mechanics is not needed, or to replace its concepts, but
rather we wish to theoretically build a machine made of springs that
mimics certain aspects of quantum theory with the hope that this
will be both of academic interest but also that it will advance the
field of phonon computation~\cite{phononics}. The main
motivation of this work was to construct the so-called tight-binding
quantum walk using a classical device (see below), the former
concept is not only relatively modern but also claimed to be useful
for quantum search algorithms~\cite{QS-1,QS-2,QS-3}.

In what follows, we first outline the relevant theory of quantum walks and provide the corresponding wave function.
Sec. III then demonstrates how to construct this wave function via
the Hilbert transform. The simulation strategy to verify our theory
and the scope and limitation of this strategy are also discussed. In
Sec. IV, the physical meaning of the wave function in classical
mechanics is analyzed, based on which, we show in Sec. V that the
wave function's modulus square can intriguingly represent the
classical energy and heat spreading densities. Finally, Sec. VI
draws our conclusion, followed by several appendices of additional
details.
\section{Theory}
\begin{figure*}
\begin{centering}
\vspace{-0.5cm} \includegraphics[width=12cm]{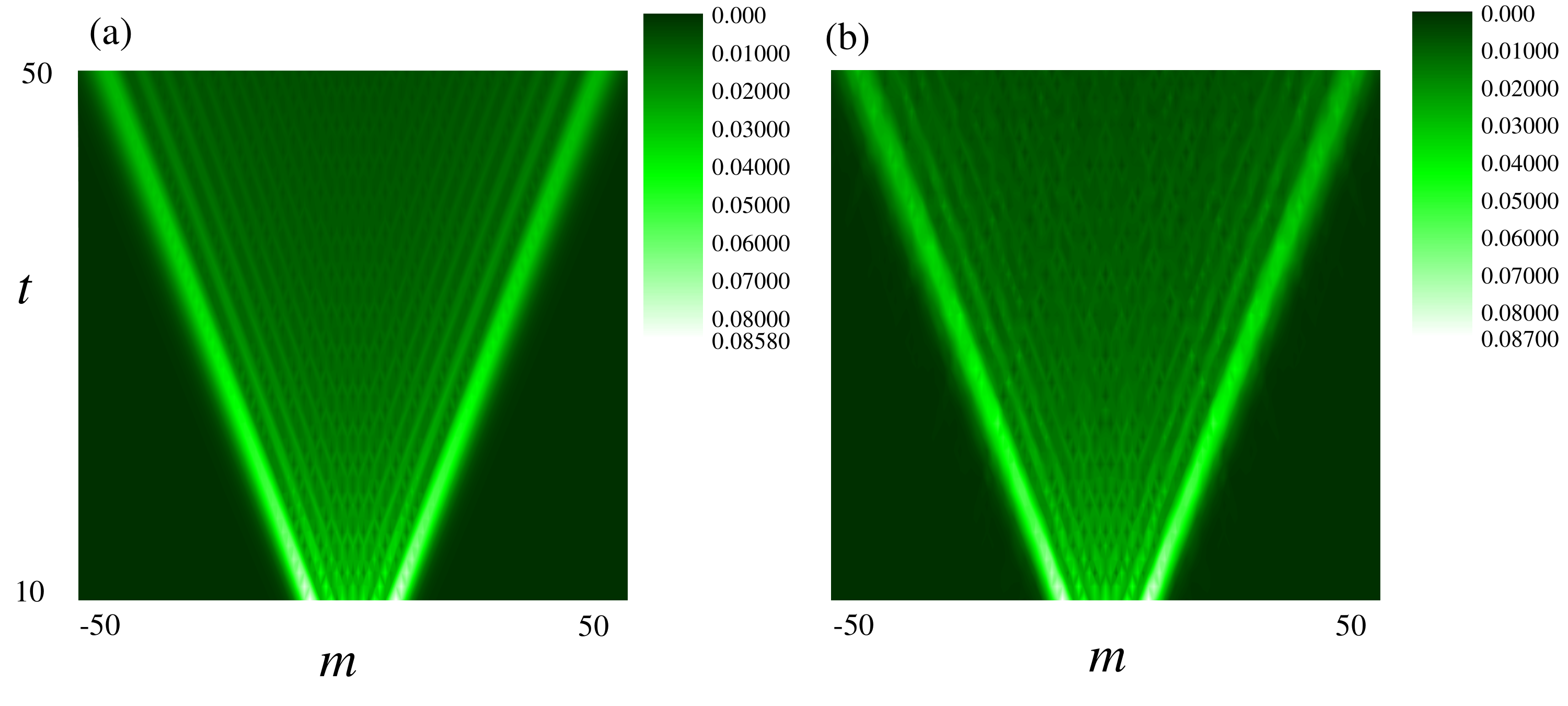} \vspace{-0.6cm}
\caption{\label{Fig1} The wave function's squared modulus
$\left|\psi_{m}(t)\right|^{2}$ from (a): the prediction of
Eqs.~(\ref{Real}-\ref{density}) with the dispersion
relation~\eqref{Dispersion} and (b): simulations using the Hilbert
transform as explained in the text, for the classical harmonic
chain. Here several short time results from $t=10$ to $t=50$ are
plotted.} \vspace{-0.6cm}
\end{centering}
\end{figure*}
At first we briefly review a particular type of quantum dynamics
that our classical device will be able to simulate. We consider a
particle on a one-dimensional (1D) lattice described by a
Hamiltonian operator $\widehat{H}$. Lattice sites $m$ are integers and the
system is infinite and translation invariant, stretching from
$-\infty $ to $\infty$. This implies that $\widehat{H}=
\sum_{j=1}^{\infty} \sum_{m=-\infty}^{\infty} \alpha_j \left
(|m\rangle \langle m+ j| + |\mit{m}\rangle \langle \mit{m}-j| \right
)$, see Refs.~\cite{QW-1,QW-2,QW-3,QW-4,QW-5,QW-6}. $\alpha_j$ are
coupling constants with dimensions of an energy. The fact that the
system is translation invariant is crucial in all of our analysis,
which is based on Fourier transform (we could treat also the system
on, say, a ring). This is also one of the main requirements from the
classical device which we will soon construct. Then for a particle
initially located on the origin, the amplitude $\psi_m(t)$ at time
$t$ is
\begin{equation} \label{WaveFunction1}
\psi_{m}(t)= \frac{1}{2 \pi} \int_{-\pi}^{\pi} e^{\rm{i}
\left(\mit{m} \mit{q}- \mit{\omega}_{\mit{q}} \mit{t} \right)}
\rm{d} \mit{q},
\end{equation}
where $q$ is the wave number, $e^{imq}$ is a free wave, and
$\omega_q$ is the frequency of a free wave, determined by
$\widehat{H} e ^{\rm{i} \mit{m} q} = \omega_q e^{\rm{i} \mit{m} q}$.
We use units where $\hbar=1$. $\omega_q$ is also called the
dispersion relation. Note that $\rm{i} \frac{\rm{d}
\mit{\psi}_m(t)}{\rm{d}\mit{t}} = \mit{\omega_q} \mit{\psi}_m(t)$,
which is easily verified from Eq.~\eqref{WaveFunction1}. The
tight-binding Hamiltonian with nearest-neighbour (NN) jumps serves
as an example. Here $\alpha_1=1$ and all other $\alpha_j=0$.
Accordingly $\rm{i} \frac{\rm{d} \psi_{\mit{m}} (\mit{t})}{\rm{d}
\mit{t}} =\psi_{\mit{m}+ \rm{1}} + \psi_{\mit{m}- \rm{1}}$.
Inserting the eigenstate solution $e^{\rm{i} \mit{m} \mit{q}}$, one
gets the dispersion relation $\omega_q=2 \cos(q)$, which in turn
bears $\psi_{m}(t)= \frac{1}{2 \pi} \int_{-\pi}^{\pi} e^{\rm{i}
\left[\mit{m} \mit{q}- \rm{2} \cos(\mit{q}) \mit{t} \right]} \rm{d}
\mit{q}= \rm{i}^{-\mit{m}} \mit{J}_{m} (\rm{2}\mit{t})$. Its modulus
square is the density $\rho
(m,t)=\left|\psi_{m}(t)\right|^{2}=\left[J_{m} (2t)\right]^{2}$,
where $J_{m}(z)$ is the Bessel function of the first
kind~\cite{QW-1,QW-6}, for a schematic presentation see
Fig.~\ref{Fig1}(a). This density has three well-known properties:
ballistic scaling $t \rho (m,t) \simeq \rho (m/t,t)$, U-like shape
and oscillations (interference) (see Appendix A). The system is also
called the tight-binding quantum walk~\cite{QW-1}; the particle's
amplitude is initially localized and may tunnel to adjacent lattice
sites. Experimentally such quantum walks can be observed e.g. in
waveguide lattices~\cite{Waveguide}.
\section{Construction of the wave function}
We next explain how to use classical particles connected
with springs to construct this wave function. In theory we provide a
classical analog-computer whose output is $\psi_m(t)$. This also
gives the exact definition and method of measurement of $\psi_m(t)$.
Before doing that, let us first divide the wave function into real
and imaginary parts. By doing so, we use $\omega_q = \omega_{-q}$ to
obtain:
\begin{equation} \label{Real}
\rm{Re} [\mit{\psi}_{m}(t)]=\frac{\rm{1}}{\rm{2} \mit{\pi}} \int_{-\pi}^{\pi} \cos (q m) \cos (\omega_q t)
\rm{d} \mit{q}
\end{equation}
and
\begin{equation} \label{Im}
\rm{Im} [\mit{\psi}_{m}(t)]=- \frac{\rm{1}}{\rm{2} \mit{\pi}} \int_{-\pi}^{\pi} \cos
(\mit{q} m) \sin (\omega_q \mit{t}) \rm{d} \mit{q}.
\end{equation}
So that the density is
\begin{equation} \label{density}
\rho (m,t)=\left|\psi_{m}(t)\right|^{2}=\{\rm{Re}[\mit{\psi}_{m}(t)]\}^{\rm{2}}+\{\rm{Im}[\mit{\psi}_{m}(t)]\}^{\rm{2}}.
\end{equation}
\subsection{The device}
Our device is a chain of classical particles whose labels are $m$,
arranged on a ring of size $N \rightarrow \infty$ with periodic
boundary conditions. The particles are interacting via linear
springs which give the phonon dispersion relation $\omega_q$ (see
examples in the Hamiltonians below). Initially, the system is in
contact with a heat bath of temperature $T$, so the initial
condition is drawn from a Boltzmann-Gibbs distribution. After
preparation, we solve the Newtonian equation of motion. The particles
evolve classically until some time $t$. We will show that the
momenta of the particles and their Hilbert transform (see definition
below) can give the wave function. For finite $T$, the classical
particles' momenta are initially random, we will employ the
correlation functions of these quantities to reproduce the quantum
dynamics. Eventually the temperature will not play any role, in the
sense that the wave function can be obtained at all temperatures.
This implies that in principle our device can operate at room
temperatures.

Like the quantum system, the main ingredient of our classical device
is that the interactions are translation invariant and linear, in
that sense all particles are identical. Essentially the label of
classical particles corresponds to the lattice site in the quantum
problem, hence our method is focused on the quantum discrete space
systems (the classical system is not space discretized). We will
consider a chain of infinite size, but all of our results can be
extended also for the finite size rings (see the discussion in Sec. III D and F).
We treat systems with NN coupling and the Hamiltonian
\begin{eqnarray} \label{Hamiltonian}
H =\sum_{m=0}^{L} \frac{p_{m}^2}{2} +V(\Delta r_m).
\end{eqnarray}
There are $N=L+1$ particles; all of them have unit mass. $p_m$ is
the momentum of the $m$th particle; $r_{m}$ is its displacement from
equilibrium position; $\Delta r_m =r_{m+1}-r_m$ denotes the NN
stretch. Applying periodic boundary conditions, we recover a ring.
More generally, the interaction potential is of the form $
\sum_{m=0}^{L} \sum_{n} \frac{1}{2} A_{n} (r_{m+n}-r_m)^2$, where
$A_n$ is the spring constant between $n$-next nearest neighbors (see
examples below).
\subsection{The real part of the wave function}
The two-body momentum correlation function is defined
as~\cite{Mazur1960}
\begin{equation} \label{MomentumC}
\rho_{p} (m,t)=\frac{\frac{1}{2} \langle p_m(t) p_0^{*}(0)+ p_0(t)
p_m^{*}(0) \rangle} {\langle |p_0(0)|^2 \rangle}
\end{equation}
with $\chi^{*}$ denoting the conjugate of $\chi$ and $\langle \cdot
\rangle$ the spatiotemporal average. This correlation function
represents the momentum correlation function of any two particles
whose separation is $m$ with a time lag $t$ since the system is
translation invariant [of course, classical momenta are real and
$p^{*}=p$ in~\eqref{MomentumC}]. Following the method proposed by
Montroll and Mazur~\cite{Mazur1960}, the
Hamiltonian~\eqref{Hamiltonian} is equivalent to $H= \frac{1}{2}
\sum_{k=0}^{L}   |P_{k}|^2 + \widetilde{\omega}_{k}^2 |R_{k}|^2 $
when applying the following normal transformation
\begin{equation} \label{Pnormal}
p_m=\sum_{k=0}^{L} C_{m,k} P_{k};
\end{equation}
\begin{equation} \label{Rnormal}
r_m=\sum_{k=0}^{L} C_{m,k} R_{k}.
\end{equation}
Here $\widetilde{\omega}_k$ is the $k$th normal mode's frequency.
For example, if one considers the harmonic chain, i.e.,
Hamiltonian~\eqref{Hamiltonian} with $V(\xi)=\xi^2/2$,
$\widetilde{\omega}_k=2 |\sin (k \pi/N)|$. $P_k$ and $R_k$ are the
normal coordinates; the matrix $C$ has the form
\begin{equation} \label{Transform}
C_{m,k}=\frac{1}{\sqrt{N}} \exp \left(2 \pi \rm{i} \frac{\mit{m} k}
{\mit{N}} \right)
\end{equation}
and satisfies
\begin{equation} \label{Transform-1}
\sum_{m=0}^{L} C_{m,k} C_{m,l}^{*}=\delta_{k,l}
\end{equation}
with $\delta$ representing the Kronecker symbol. So, for each normal
mode, we obtain the evolution equation
\begin{equation}
\frac{\rm{d}^2 \mit{R_k}} {\rm{d} \mit{t}^{\rm{2}}}+
\widetilde{\omega}_k^2 R_k=0,
\end{equation}
which determines $R_k$ and $P_k$:
\begin{equation} \label{RnormalT}
R_k(t)=[P_k(0)/\widetilde{\omega}_k] \sin (\widetilde{\omega}_k
t)+R_k(0) \cos (\widetilde{\omega}_k t)
\end{equation}
and
\begin{equation} \label{PnormalT}
P_k(t)=P_k(0) \cos (\widetilde{\omega}_k t) - \widetilde{\omega}_k
R_k(0) \sin (\widetilde{\omega}_k t).
\end{equation}
Substitute Eqs.~(\ref{Pnormal}-\ref{PnormalT})
into~\eqref{MomentumC} and use the following equipartition
conditions: $\langle P_k(0) P_l^{*} (0)\rangle=k_B T \delta_{k,l}$;
$\langle R_k(0) R_l^{*} (0)\rangle=  k_B T \delta_{k,l}/
\widetilde{\omega}_k^2$, and $\langle R_k(0) P_l^{*} (0)\rangle=0$
($k_B$ is the Boltzmann constant), one finally
obtains~\cite{Mazur1960}
\begin{equation}
\rho_{p} (m,t)=\frac{1}{N} \sum_{k=0}^{L} \cos\left(\frac{2 \pi m
k}{N}\right) \cos (\widetilde{\omega}_k t).
\end{equation}
Taking $N$ (or $L$) $\rightarrow \infty$ and accordingly $2 \pi k/N
\rightarrow q$, we have $\widetilde{\omega}_k \rightarrow \omega_q$.
In particular we obtain the classical dispersion relation of the
harmonic chain as a specific example:
\begin{equation} \label{Dispersion}
\omega_q=2|\sin(q/2)|.
\end{equation}
For any classical chain's dispersion relation we have:
\begin{equation} \label{PredictionM}
\rho_{p} (m,t)=\frac{1}{2 \pi} \int_{- \pi}^{\pi} \cos (q m) \cos
(\omega_q t) \rm{d} \mit{q},
\end{equation}
which is just $\rm{Re} [\mit{\psi}_{m}(t)]$ from~\eqref{Real}. Note
that the classical dispersion of a system of springs is controlled
by the interactions of a bead with its neighbors. If these
interactions are controllable, we may obtain rather general forms of
the classical dispersion, as we demonstrate in examples below.
\subsection{Imaginary part of $\psi_m(t)$: the $\pi/2$-shifted momentum correlation function}
Define the following cross-correlation function
\begin{equation} \label{PPmomentumC}
\rho_{\widetilde{p}}(m,t)=\frac{1}{2} \frac{\langle \widetilde{p}_m(t)
p_0^{*}(0)+ \widetilde{p}_0(t) p_m^{*}(0) \rangle} {\langle |p_0(0)|^2
\rangle},
\end{equation}
where $\widetilde{p}_m(t)$ is what we call the $\pi/2$-shifted momentum. Since $p_m(t)$ is a linear combination of normal modes, $\widetilde{p}_m(t)$ can be obtained by shifting the underlying normal modes
\begin{eqnarray} \label{PPnormalT}
\widetilde{P}_k(t) &= &P_k(0) \cos (\widetilde{\omega}_k t + \pi/2) -
\widetilde{\omega}_k R_k(0) \sin (\widetilde{\omega}_k t +\pi/2) \nonumber \\
&= &-P_k(0)
\sin (\widetilde{\omega}_k t) -\widetilde{\omega}_k R_k(0) \cos
(\widetilde{\omega}_k t)
\end{eqnarray}
and performing the transformation $\widetilde{p}_m (t)=\sum_{k=0}^{L} C_{m,k} \widetilde{P}_{k} (t)$ [see Eq.~\eqref{Pnormal}]. Now use~\eqref{PPnormalT} and repeat the steps from $\rho_{p} (m,t)$ to $\rho_{\widetilde{p}}(m,t)$. One gets
\begin{equation}
\rho_{\widetilde{p}}(m,t)= -\frac{1}{N} \sum_{k=0}^{L} \cos \left(\frac{2 \pi m k}{N}\right) \sin (\widetilde{\omega}_k t).
\end{equation}
Finally take $N\rightarrow \infty$, to obtain
\begin{equation}
\rho_{\widetilde{p}}
(m,t)= - \frac{1}{2 \pi} \int_{-\pi}^{\pi} \cos (\mit{q} m) \sin
(\omega_q \mit{t}) \rm{d} \mit{q},
\end{equation}
which is $\rm{Im} [\mit{\psi}_{m}(t)]$ in~\eqref{Im}.
\subsection{Hilbert transform}
The $\pi/2$-shifted momentum is mathematically equivalent to the
negative Hilbert transform~\cite{Hilbert} of the momentum. In fact,
the Hilbert transform's effect is a $\pi/2$ phase shift of each
frequency components. For an arbitrary $p_m(t)$, one defines its
Hilbert transform $\mathscr{H}[p_m(t)]$ as (see Ref.~\cite{Hilbert})
\begin{equation}
\mathscr{H}[p_m(t)]=\frac{1}{\pi}
\rm{p}.\rm{v}.\int_{-\infty}^{\infty}
\frac{\mit{p_m}(\tau)}{\mit{t}-\tau} \rm{d} \mit{\tau},
\end{equation}
where $\rm{p}.\rm{v}.$ indicates the Cauchy principal value. This
definition bears $\mathscr{H}[\sin (\widetilde{\omega}_k t)]=-\cos
(\widetilde{\omega}_k t)$ and $\mathscr{H}[\cos
(\widetilde{\omega}_k t)]=\sin (\widetilde{\omega}_k t)$;
consequently $\widetilde{p}_m(t) = -\mathscr{H}[p_m(t)]$ . With this
definition, $p_m(t)$ and $\mathscr{H}[p_m(t)] = -
\widetilde{p}_m(t)$ form a complex conjugate pair that defines the
so-called analytic signal
\begin{equation}
Z_m(t) = p_m(t) - \rm{i} \mit{\mathscr{H}} [p_m(t)].
\end{equation}
Applying Fourier transform $\widehat{Z}_k(t) = \sum_{m=0}^{L}
C^{*}_{m,k} \mit{Z}_m(t)$, we have
\begin{equation} \label{Temp1}
\widehat{Z}_k(t) = P_k(t) - \rm{i} \mit{\mathscr{H}} [P_k(t)]=P_k(t)
- \rm{i} \widetilde{\omega}_{\mit{k}} \mit{R}_k(t).
\end{equation}
The last term of~\eqref{Temp1} arises because
$\mathscr{H}[P_k(t)]=-\widetilde{P}_k(t)= \widetilde{\omega}_k
R_k(t)$ in view of~\eqref{RnormalT} and~\eqref{PPnormalT}. We now
conjecture that the normalized correlation function $\psi_m(t) =
\frac{\langle Z_m(t) Z_{\rm{0}}^{*}(\rm{0}) \rangle}{\langle
Z_{\rm{0}}(\rm{0}) \mit{Z}_{\rm{0}}^{*}(\rm{0}) \rangle}$ is the
wave function. This is best seen from its Fourier transform
$\widehat{\psi}_k(t) = \sum_{m=0}^{L} C^*_{m,k} \psi_m(t)$. We have
\begin{eqnarray} \label{schroedinger}
\rm{i} \frac{\rm{d} \mit{\widehat{\psi}}_k(t)}{\rm{d}\mit{t}} &=& \rm{i} \frac{1}{\langle \mit{Z}_{\rm{0}}(\rm{0}) \mit{Z}_{\rm{0}}^{*}(\rm{0}) \rangle} \left\langle \mit{Z}_{\rm{0}}^{*}(\rm{0}) \frac{\rm{d} \mit{\widehat{Z}}_k(t) } {\rm{d}\mit{t}} \right\rangle\nonumber\\
&=& \rm{i} \frac{1}{\langle \mit{Z}_{\rm{0}}(\rm{0}) \mit{Z}_{\rm{0}}^{*}(\rm{0}) \rangle} \left\langle \mit{Z}_{\rm{0}}^{*}(\rm{0}) \left[\frac{\rm{d} \mit{P}_k(t) } {\rm{d}\mit{t}} - \rm{i} \widetilde{\omega}_{\mit{k}} \frac{\rm{d} \mit{R}_k(t) } {\rm{d}\mit{t}} \right] \right\rangle \nonumber \\
&=& \rm{i} \frac{1}{\langle \mit{Z}_{\rm{0}}(\rm{0}) \mit{Z}_{\rm{0}}^{*}(\rm{0}) \rangle} \left\langle \mit{Z}_{\rm{0}}^{*}(\rm{0}) \left[- \widetilde{\omega}^{2}_{\mit{k}} \mit{R}_k(t) - \rm{i} \widetilde{\omega}_{\mit{k}} \mit{P}_{k}(t) \right] \right\rangle \nonumber \\
&=& -\rm{i}^{2} \frac{1}{\langle \mit{Z}_{\rm{0}}(\rm{0}) \mit{Z}_{\rm{0}}^{*}(\rm{0}) \rangle} \widetilde{\omega}_{\mit{k}} \left\langle \mit{Z}_{\rm{0}}^{*}(\rm{0}) \left[ \mit{P}_k(t) - \rm{i} \widetilde{\omega}_{\mit{k}} \mit{R}_k(t) \right] \right\rangle \nonumber \\
&=& \widetilde{\omega}_{\mit{k}} \left\langle
\frac{\mit{Z}_{\rm{0}}^{*}(\rm{0})
\mit{\widehat{Z}}_k(t)}{\mit{Z}_{\rm{0}}(\rm{0})
\mit{Z}_{\rm{0}}^{*}(\rm{0})} \right\rangle =
\widetilde{\omega}_{\mit{k}} \mit{\widehat{\psi}}_k(t),
\end{eqnarray}
which is just the Schr\"{o}dinger equation in Fourier space that we
mentioned at the beginning. $\frac{\rm{d} \mit{R}_k(t) }
{\rm{d}\mit{t}}$ and $\frac{\rm{d} \mit{P}_k(t) } {\rm{d}\mit{t}}$
were obtained from~\eqref{RnormalT} and~\eqref{PnormalT}. Turning
back to lattice space, we get the wave function's real and imaginary
parts. One might think that the relation between $\psi_m$ and $Z_m$
is just a curious mathematical coincidence. However, the quantum and
the classical systems share wave-like properties. For this reason,
we searched for this intriguing exact correspondence between the
classical and quantum worlds. The appearance of the Schr\"{o}dinger
equation confirms our idea that classical systems' correlation
functions can be used to obtain quantum dynamics. It is worthwhile
to note that the theory works also for finite number of particles.
Translation invariance of both, the classical and corresponding
quantum system, ensures the applicability of Fourier analysis. Thus,
the limit of $N \rightarrow \infty$ is not a general request.
\subsection{Construction of the wave function by simulations}
In Fig.~\ref{Fig1} we compared the wave function obtained from a
quantum walk with the one obtained from a harmonic chain. To do so,
we first computed the wave function's modulus square $|\psi_m(t)|^2$
by plugging the dispersion relation~\eqref{Dispersion} into
Eqs.~(\ref{Real}-\ref{density}), see Fig.~\ref{Fig1}(a). Then we
simulated a harmonic chain with $N=4001$ particles, initially
connected to a heat bath with $T = 0.5$ (see simulation detail in
Appendix B). The same values are used below. The results are also
verified for other temperatures. Employing the correlation
functions, $\rho_{p}(m,t)=\frac{\langle p_{i+m}(t) p_{i}(0)
\rangle}{\langle p_{i}(0) p_{i}(0) \rangle}$ and
$\rho_{\widetilde{p}}(m,t)=\frac{\langle \widetilde{p}_{i+m}(t)
p_{i}(0) \rangle}{\langle  p_{i}(0)  p_{i}(0) \rangle}$, the wave
function's modulus square is computed via $|\psi_m(t)|^2 =
[\rho_{p}(m,t)]^2+ [\rho_{\widetilde{p}}(m,t)]^2$. The result is
depicted in Fig.~\ref{Fig1}(b). In practice we apply the Hilbert
transform on each particle's momentum to obtain
$\rho_{\widetilde{p}}(m,t)$; the details are also described in
Appendix B.
\subsection{Scope and limitation}
We focus on simple systems with translational invariance, both for
the quantum dynamics and the classical device. The classical
particles are linearly coupled. Their label $m$ corresponds to a
lattice site of the quantum system. The linear chain describes a
single quantum particle on a lattice. It can be a finite ring of
elements, or stretch to infinity. Furthermore, the initial state of
the classical system is a thermal one. This implies uncorrelated
momenta. At $t=0$, the momentum correlation function is a Kronecker
delta, consequently the quantum particle is initially localized.
These are certainly constraints on the generality of our device and
we are still far from a universal classical machine capable of
simulating all aspects of quantum mechanics. Still the Hilbert
transform technique is encouraging and hopefully further research
will unravel more general devices. For that reason we proceed to
show that the wave function can, in the classical context, be used
to describe the kinetic energy, stretch, and total energy and heat
correlation functions.
\section{The physical meaning of the wave function in classical mechanics}
So far, we have shown how to construct the wave function from a
linear chain's correlation functions. Now we would like to point out
the physical meaning of the wave function in the classical domain.
It is a physically significant observable describing various
equilibrium correlation functions beyond the momentum and its
Hilbert transformed correlation functions (see Appendix C). First,
one can rigorously prove that the square of the wave function's real
part $\left\{\rm{Re}[\psi_{\mit{m}}(\mit{t})]\right\}^2$ is the
kinetic energy correlation function (see Appendix C1). Second, the
non-normalized stretch correlation function $C_{\Delta r } (m,t)$,
defined by $\langle \Delta r_m(t) \Delta r_0(0) \rangle$, is related
to $\psi_{m}(t)$ (see Appendix C2)
\begin{equation}
\frac{\rm{d}^{\rm{2}}}{\rm{d}\mit{t}^{\rm{2}}} \left[
\frac{C_{\Delta r} (m,t)} {k_B T}\right]=\rm{Re} \left
[\psi_{\mit{m}+\rm{1}} (\mit{t}) + \psi_{\mit{m}-\rm{1}} (\mit{t}) -
\rm{2} \psi_{\mit{m}} (\mit{t}) \right].
\end{equation}
Similarly the stretch-momentum cross-correlation function $C_{\Delta
r p} (m,t)$, defined by $\langle \Delta r_m(t) p_0(0) \rangle$, is
shown to be related to $\psi_{m}(t)$ by (see Appendix C3)
\begin{equation}
\frac{\rm{d}}{\rm{d}\mit{t}} \left[ \frac{C_{\Delta r p} (m,t)} {k_B
T}\right]=\rm{Re} \left [ \psi_{\mit{m}+ \rm{1}}
(\mit{t})-\psi_{\mit{m}} (\mit{t}) \right ],
\end{equation}
which gives
\begin{align}
    & \nonumber
        C_{\Delta r p} (m,t)=\frac{k_B T}{2 \pi}  \times
    \\ &
        \int_{-\pi}^{\pi} \frac{
\rm{sin} (\mit{\omega_q t})} {\omega_q} \left[\rm{cos} \mit{(q m+ q
)} - \rm{cos} (\mit{q m}) \right] \rm{d} \mit{q}.
\end{align}

Furthermore, we will demonstrate that the potential energy and the
total energy correlation functions can also be related to the wave
function (see Appendix C4). Based on all of these facts, below we
will provide evidences that the density $\rho(m,t)=|\psi_m(t)|^2$
describes both the normalized (total) energy and heat correlation
functions, in the long time limit~\cite{Dharnew}. Thus, we see, that
the proposed wave function contains rich physical information on the
classical chain.
\begin{figure}
\begin{centering}
\vspace{-.6cm} \includegraphics[width=6.6cm]{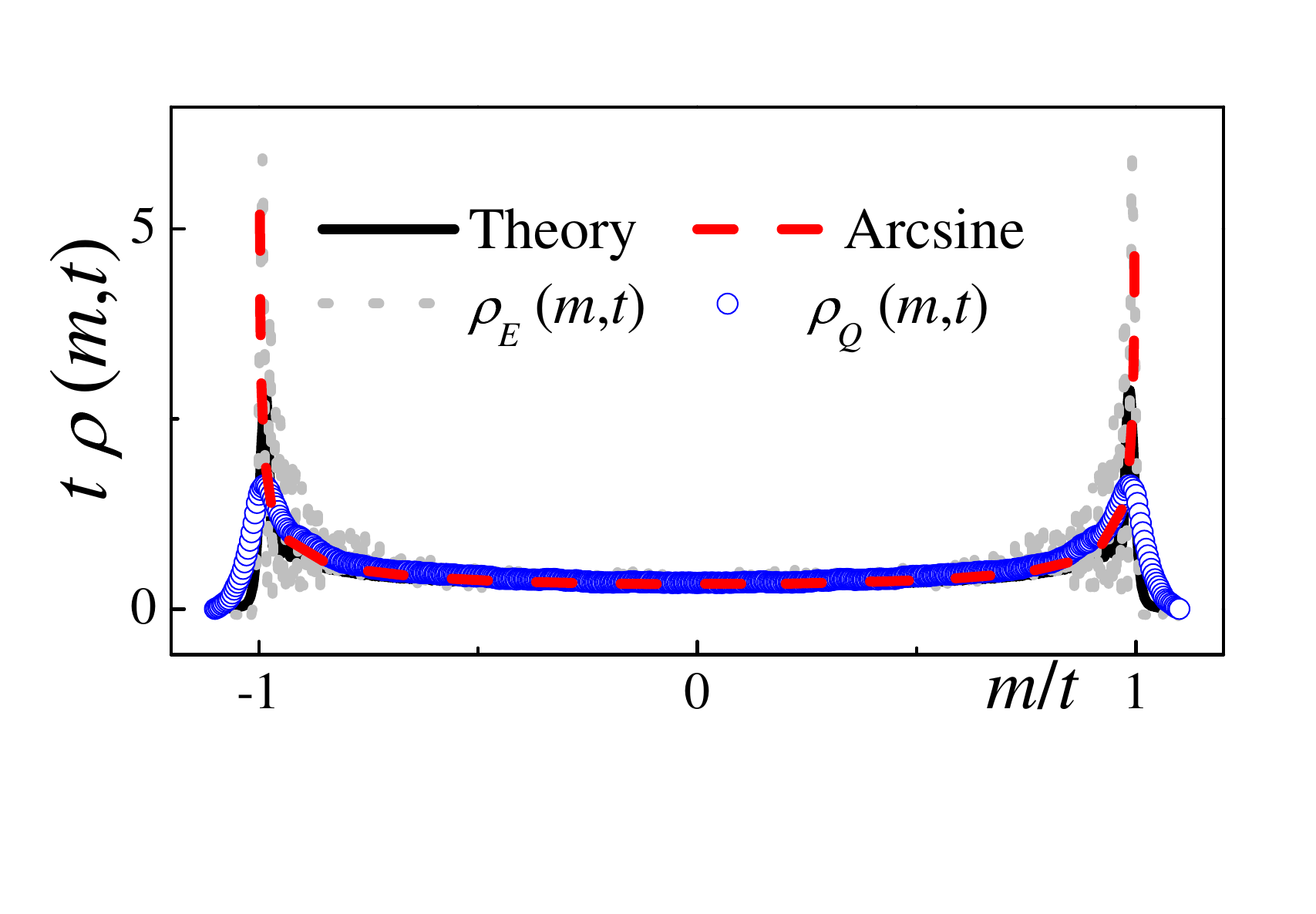} \vspace{-1.2cm}
\caption{\label{Fig2} The rescaled densities for the harmonic chain
($t=600$), obtained from plugging Eq.~\eqref{Dispersion} into
Eqs.~(\ref{Real}-\ref{density}), are compared with the prediction
from the Arcsine distribution and simulations.} \vspace{-0.6cm}
\end{centering}
\end{figure}
\section{Energy and heat spreading densities}
\begin{figure*}
\begin{centering}
\vspace{-.83cm} \includegraphics[width=15cm]{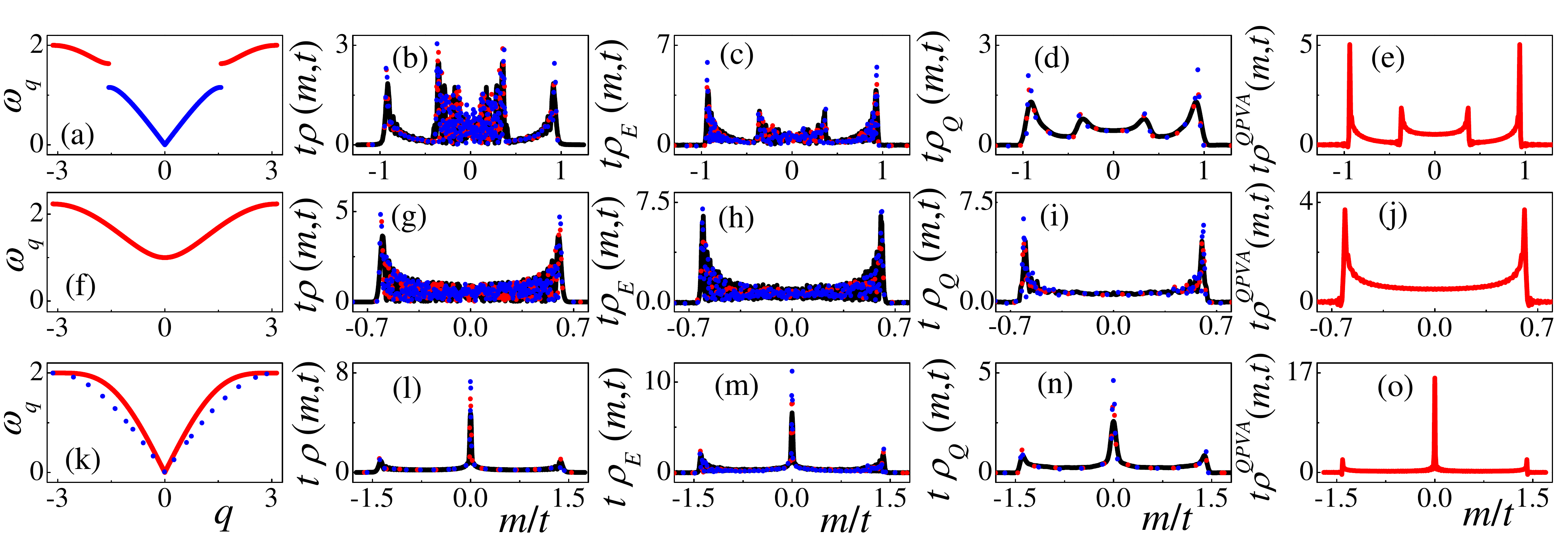}
\vspace{-.5cm} \caption{\label{Fig3} Dispersion relations [(a),
(f), (k)], rescaled densities from the predictions of
Eqs.~(\ref{Real}-\ref{density}) with $\omega_q$ in Table~\ref{T1}
[(b), (g) and (l)] and simulations [$\rho_E(m,t)$: (c), (h) and (m);
$\rho_Q(m,t)$: (d), (i) and (n)], and the predicted densities from
QPVA [(e), (j), (o)] for Model II (a)-(e); Model III (f)-(j), and
Model IV (k)-(o), respectively. For all the densities, three long
times' results [solid ($t=200$), dashed ($t=400$) and dotted
($t=600$)] are rescaled for comparison. In (k) the harmonic chain's
dispersion relation (dashed) is plotted for comparison. Note that
$\rho_Q (m,t)$ and $\rho_E(m,t)$ are not mathematically identical,
though both are well approximated by $\rho(m,t)$. } \vspace{-.6cm}
\end{centering}
\end{figure*}
\subsection{U shaped density for harmonic chain}
As demonstrated in Fig.~\ref{Fig1}, given a phonon dispersion
relation, one can simulate a quantum walk by observing the wave
function's modulus square [see Eq.~\eqref{density}]. Here we compare
the prediction of Eqs.~(\ref{Real}-\ref{density}) to the simulations
of energy and heat spreading densities for a harmonic chain in the
long-time limit. The energy and heat spreading densities are usually
obtained from the following correlation functions, i.e., $\rho_{E}
(m,t)=\frac{\langle \Delta E_{i+m}(t) \Delta E_{i}(0)
\rangle}{\langle \Delta E_{i}(0) \Delta E_{i}(0) \rangle}$ and
$\rho_{Q} (m,t)=\frac{\langle \Delta Q_{i+m}(t) \Delta Q_{i}(0)
\rangle}{\langle \Delta Q_{i}(0) \Delta Q_{i}(0) \rangle}$, where
$E_{i}(t)$ and $Q_{i}(t)$ are the energy and heat densities at
location $i$ and time $t$. $\Delta E_{i}(t)$ and $\Delta Q_{i}(t)$
are their fluctuations, respectively (see
Refs.~\cite{Forster,Liquid,Zhao2006,Chen2013,Xiong2016-1,Xiong2016-2}
and Appendix B for detailed definitions).

Intriguingly, we find that both densities nicely match and converge
to an U-shape (see Fig.~\ref{Fig2}). This U-shaped solution is
L\'{e}vy's well-known Arcsine law that also describes the occupation
times of an 1D Brownian particle in a half space
(see~\cite{LevyArcsine,RW-3,RW-4}).
\subsection{Quasi particle velocity approach}
To understand this U-shape, we here propose a method we call ``quasi
particle velocity approach'' (QPVA), which perfectly gives rise to
the Arcsine law in the harmonic chain.

Our main idea is that in a linear chain, phonons can be understood as
quasi particles. They start on the origin and will spread out
ballistically with $m=vt$. In this case, given the probability
distribution function (PDF) of the particle's velocity, $h(v)$, the
density is given by
\begin{equation} \label{Stemp2}
    \rho (m,t)= h(v) \left.\frac{\rm{d}\mit{v}}{\rm{d}\mit{m}}\right|_{v = m/t} = \frac{1}{t} h(\tfrac{m}{t}).
\end{equation}
To obtain the velocity PDF, one first takes the Fourier transform of
$h(v)$
\begin{equation}
    \widetilde{h}(\mu)=\int_{-\infty}^{\infty} e^{\rm{i} \mu \mit{v}} h(v) \rm{d} \mit{v} = \langle e^{i\mu v} \rangle
\end{equation}
which is just the characteristic function. $h(v)$ is obtained by
inverse Fourier transform of this characteristic function. Thus, we
need to find $\left \langle e^{\rm{i} \mu \mit{v}}\right\rangle$. In
our study, we consider highly localized initial conditions. This
corresponds to an uniform distribution of wave vectors, $q$. A free
wave's velocity $v_q$ can be identified with the group velocity
$(\mathrm{d} \omega_q / \mathrm{d} q)$. Hence in the harmonic chain
with dispersion relation $\omega_{q} =2 \left|\sin
\left(q/2\right)\right|$ we have
\begin{eqnarray}
v_{q}=
\begin{cases}
\cos(\frac{q}{2}), &q\geq 0\cr -\cos(\frac{q}{2}), &q<0
\end{cases};
\end{eqnarray}
then
\begin{eqnarray}
\left \langle e^{\rm{i} \mu \mit{v}}\right\rangle &=& \frac{1}{2\pi}
\int_{-\pi}^{\pi} e^{\rm{i} \mu \mit{v}_{q}} \rm{d} \mit{q} \nonumber \\
&=& \frac{1}{2\pi} \left[\int_{0}^{\pi} e^{\rm{i} \mu
\cos(\frac{\mit{q}}{\rm{2}})} \rm{d}
\mit{q} + \int_{-\pi}^{\rm{0}} e^{-\rm{i} \mu \cos(\frac{\mit{q}}{\rm{2}})} \rm{d} \mit{q} \right] \nonumber \\
&=& J_{0}(\mu),
\end{eqnarray}
hence
\begin{equation} \label{Stemp1}
h(v)=\frac{1}{2 \pi} \int_{-\infty}^{\infty} e^{-\rm{i} \mu \mit{v}}
J_{0}(\mu) \rm{d} \mit{\mu}=\frac{\rm{1}}{\pi
\sqrt{\rm{1}-\mit{v}^{\rm{2}}}}.
\end{equation}
Finally, substituting~\eqref{Stemp1} into~\eqref{Stemp2}, we finally
get the rescaled Arcsine distribution
\begin{equation}
\rho (m,t)=h(v)
\left.\frac{\rm{d}\mit{v}}{\rm{d}\mit{m}}\right|_{v=m/t}=\frac{1}{t}
\frac{1}{\pi \sqrt{1-(m/t)^{2}}},
\end{equation}
which has the predicted U-shape and the correct ballistic scaling in
the long time limit.
\subsection{Non-universal shapes: dependent on $\omega_q$}
To demonstrate that the equivalence between $\rho_E(m,t)$,
$\rho_Q(m,t)$ and Eqs.~(\ref{Real}-\ref{density}) holds in general,
we consider three more complicated Hamiltonians with different
dispersion relations. Model II: a chain with alternating coupling
including two branches of phonons, $H= \sum_{m=0}^{L/2}
\left(p_{2m-1}^2+ p_{2m}^2\right)/2+k_1 V(\Delta r_{2m}) +k_2
V(\Delta r_{2m+1})$ (see Ref.~\cite{Alternating-Coupling}), where
$\Delta r_{2m}=r_{2m}-r_{2m-1}$ and $V(\xi)=\xi^2/2$ (the same
below). Model III: the lattice $\phi^4$ system but with linear
on-site potential $H=\sum_{m=0}^{L} p_m^2/2 +V(\Delta r_m) +
r_m^2/2$ (see Ref.~\cite{Chaos}). And finally Model IV: a 1D lattice
with next-nearest-neighbour (NNN) coupling $H= \sum_{m=0}^{L}
p_{m}^2/2+V(r_{m+1}-r_{m})+\gamma V(r_{m+2}-r_{m})$ (see
Refs.~\cite{NNN-1,NNN-2}). The particular phonon dispersion
relations are listed in Table~\ref{T1} and also plotted in
Fig.~\ref{Fig3}(a,f,k).
\begin{table}[!hbp]
\begin{centering}
\vspace{-.1cm}
\begin{tabular}{c c}
 \hline
  Models  & Dispersion relation \\ \hline
  II &  $\omega_{q}^{\pm}=\sqrt{k_1+k_2 \pm \sqrt{k_1^{2}+k_2^{2}+2 k_1 k_2 \cos (2q)}}$ \\
  III   & $\omega_{q} =\sqrt{4 \sin^2\left(q/2\right)+1}$  \\
  IV   &  $\omega_{q} =2 \sqrt{\sin^2\left(q/2\right)+\gamma \sin^2\left( q \right)}$ \\
 \hline
 \vspace{-.2cm}
\end{tabular}
\caption{\label{T1} Phonon dispersion relations for Model II-IV, where $\omega_{q}^{-}$ ($\omega_{q}^{+}$) denotes the frequency of acoustic (optical) phonons; $k_1$ and $k_2$ are the strength of the adjacent couplings; $\gamma$ represents the comparative strength of the NNN to the NN couplings. We use $k_1 =1/3$, $k_2 = 2/3$ and set $\gamma=0.25$ here.} \vspace{-.3cm}
\end{centering}
\end{table}

Figure~\ref{Fig3}(b,g,l) show the prediction for $\rho_E(m,t)$ and
$\rho_Q(m,t)$ from Eqs.~(\ref{Real}-\ref{density}) and
Fig.~\ref{Fig3}(c,h,m,d,i,n) depicts the simulated values. All
features of the simulation are nicely captured by our prediction.
This clearly demonstrates the generality of our approach. We note
that extending our theory to Model III (IV) is straightforward, one
just needs to integrate Eqs.~(\ref{Real}-\ref{density}) with the
$\omega_q$ given in Table~\ref{T1}. However, for Model II, one
should consider contributions from both acoustic $\psi_{m}^{-}(t)=
\frac{1}{2 \pi} \int_{-\pi/2}^{\pi/2} e^{\rm{i} \left(\mit{m}
\mit{q}-\omega_{q}^{-} \mit{t} \right)} \rm{d} \mit{q}$ and optical
phonons $\psi_{m}^{+}(t)=\frac{1}{2 \pi} \left[
\int_{\pi/2}^{\pi}e^{\rm{i} \left(\mit{m} \mit{q}- \omega_{q}^{+}
\mit{t} \right)} \rm{d} \mit{q} + \int_{-\pi}^{-\pi/\rm{2}}
e^{\rm{i}\left(\mit{m} \mit{q}- \omega_{q}^{+} \mit{t} \right)}
\rm{d} \mit{q} \right]$. The solution then is
$\rho(m,t)=\left|\psi_{m}^{-}(t)+ \psi_{m}^{+}(t)\right|^{2}$ (see
detailed analysis in Appendix D). The excellent agreement between
simulation and theory indicates that our theory also works well for
systems with two branches of phonons. This unusual fact may
stimulate the conception of new phononics devices~\cite{Note}, since
one may be able to identify independently the contributions of
acoustic and optical phonons.
\begin{figure*}
\begin{centering}
\vspace{-.3cm} \includegraphics[width=14cm]{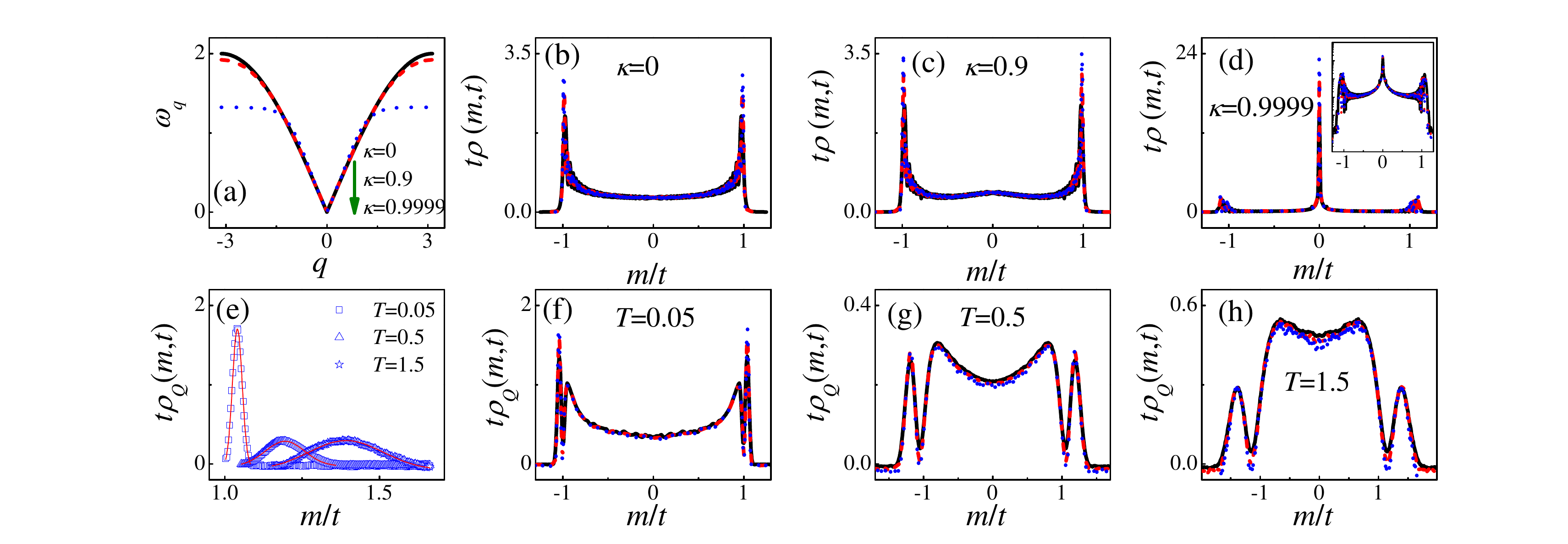} \vspace{-0.45cm}
\caption{\label{Fig4} Dispersion relations (a), rescaled densities
[solid ($t=200$), dashed ($t=400$) and dotted ($t=600$)] for the
Toda chain, from the predictions of Eqs.~(\ref{Real}-\ref{density})
[(b)-(d)], and simulations [(f)-(h)]. (e): the side peaks in (f)-(h)
are fitted by Gaussian distributions $N (\nu, \sigma^{2})$ with
means $\nu$ and variances $\sigma^{2}$: $T=0.05$ $(1.04,0.00025)$;
$T=0.5$ $(1.19,0.004)$ and $T=1.5$ $(1.39,0.015)$. In the inset of
(d) the y-axis is logarithmic.} \vspace{-.6cm}
\end{centering}
\end{figure*}

From Fig.~\ref{Fig3} one sees that the ballistic transport exhibits
non-universal features (unlike the super-diffusion investigated for
example in~\cite{RW-5}). Motivated by this, we use the
proposed QPVA to find the long time asymptotic behaviour of the
packet, and want to understand how exactly it depends on $\omega_q$.
Similar to the harmonic chain, one may view the problem as a packet
of quasi particles initially localized in space. Each of the quasi
particles travels with speed $v_{q}=\frac{\rm{d}
\omega_{\mit{q}}}{\rm{d} \mit{q}}$. All values of $-\pi<q<\pi$ are
uniformly distributed. We find
\begin{equation} \label{QPVA}
t \rho (m,t) \sim h(v) = \frac{1}{2 \pi} \int_{-\infty}^{\infty}
e^{-\rm{i} \mu \mit{v}} \left( \frac{1}{2\pi} \int_{-\pi}^{\pi}
e^{\rm{i} \mu \mit{v}_{q}} \rm{d} \mit{q} \right) \rm{d} \mit{\mu}
\end{equation}
with $v=m/t$. Hence $t\rho(m,t)$ is the inverse Fourier transform
$h(v)$ of $\frac{1}{2\pi} \int_{-\pi}^{\pi} e^{\rm{i}
\mu\mit{v}_{q}} \rm{d} \mit{q}$, $\mu\leftrightarrow v$. For the
harmonic chain with the dispersion relation~\eqref{Dispersion}, this
yields the mentioned Arcsine law. For other $\omega_q$,
Fig.~\ref{Fig3} shows excellent agreement between simulations and
our formula. We note that this picture only yields the long time
limit of $\rho(m,t)$. It does not feature the fine oscillations so
typical of the ballistic dynamics (These oscillations are presented
in Fig.~\ref{Fig1}, and they are important in the short time limit).
\subsection{Discussion on non-linear integrable Toda system}
As a last example, we consider the celebrated Toda chain with
Hamiltonian~\eqref{Hamiltonian} and $V(\xi)=a {\rm{e}}^{-\mit{b}
\xi}/b+a \xi + c$, which also bears ballistic
transport~\cite{Toda1979}. The aim here is to discuss whether our
ideas could apply to a non-linear integrable system. The Toda
chain's dispersion relation has the explicit form $\omega_{q}=
\frac{\pi}{K_{1}} / \sqrt{\frac{1}{{\rm{Sn}}^2(K_{1} q / \pi)} -1 +
\frac{K_{2}}{K_{1}}}$ (see Ref.~\cite{Toda1979}). Inserting this
relation into Eqs.~(\ref{Real}-\ref{density}) gives predictions for
the packet shape, which can be compared with simulations. Here
${\rm{Sn}}$ is the Jacobian elliptic function with modulus $\kappa$.
$\kappa$ is some constant determined by $a$, $b$ and $T$ that
describes the non-linearity, $0 \le \kappa < 1$. $K_{1}(\kappa)$
[$K_{2}(\kappa)$] is the complete elliptic integral of the first
(second) kind. In contrast to the linear systems, $\omega_{q}$ is no
longer a single line, it rather depends on $\kappa$, and hence on
$T$. For $\kappa$ close to zero, we get harmonic behaviour. As
$\kappa \rightarrow 1$, $v_{q}$ (the group velocity defined by
$\frac{\rm{d} \omega_{\mit{q}}}{\rm{d} \mit{q}}$) vanishes for large
$q$ and increases for small $q$ [see Fig.~\ref{Fig4}(a)]. We show
below how these features affect the density's shape.

Predictions for the densities are shown in Fig.~\ref{Fig4}(b)-(d)
for several values of $\kappa$. For small $\kappa$, the Toda chain
behaves like a linear system: an U-shape can be observed
[Fig.~\ref{Fig4}(b)]. As $\kappa$ increases, the density's central
parts become humped together and the two side peaks emerge at
$|m/t|>1$ [see Fig.~\ref{Fig4}(c)-(d) and inset].

Numerical results for $\rho_{Q} (m,t)$ are given for comparison. The
simulations are performed with $ V(\xi)= {\rm{e}}^{-\xi} +\xi -
\rm{1}$, i.e. $a=b=-c=1$ (see details in Appendix B). Since we lack
information about the $T$-dependence of the dispersion relation, the
comparison is just indirect and we are unable to provide predictions
for a given $T$. From Fig.~\ref{Fig4}(f)-(h) we can indeed verify
the two trends of $\rho (m,t)$: for small $T$, the central parts of
$\rho_{Q} (m,t)$ are very similar to the U-shape and slight side
peaks at $|m/t| > 1$ can already be identified [Fig.~\ref{Fig4}(f)].
As $T$ increases, more and more front parts emerge and also a hump
in the central part appears [Fig.~\ref{Fig4}(g)-(h)]. Evidently the
trends in simulations and predictions coincide, although both do not
match precisely.

We also examined the front parts located at $|m/t|
> 1$ and find that they can be fitted quite well with a Gaussian
distribution [Fig.~\ref{Fig4}(e)]. As $T$ increases, so do the mean
and variance of the Gaussian. This is in good agreement with the
velocity fluctuations conjecture that was suggested by the L\'evy
walks approach for predicting the non-linear non-integrable
Fermi-Pasta-Ulam-$\beta$ chain's density~\cite{RW-5}.
\section{Conclusion}
In summary, we have demonstrated how to use classical chains of
springs to simulate quantum dynamics (in particular, the quantum
walk). To do this, we have suggested to use the momenta's and their
Hilbert transform's pair correlation function to construct a quantum
like wave function. Such a strategy successfully solves the
challenges addressed by Feynman on this topic. Therefore it provokes
the general idea of making a classical machine to reproduce quantum
dynamics. We leave to future work if the Hilbert transform technique
can be used to model other aspects of quantum mechanics, e.g., spin,
magnetic field, and many-body systems.
We will show in a future publication that our device can work also for non-translation invariant systems and can thus model transport in e.g. disordered systems.
In that case the Fourier analysis used all along this text does not work, and other tools are needed to solve the problem.

Previous methods~\cite{Bohm-1,Bohm-2,Nelson} used the solution of the
Schr\"odinger equation $\psi= R \rm{e}^{\rm{i} \mit{S}}$ and then
constructed a classical ensemble that yields back $\psi$. Their
paths are generated using a deterministic law $\dot{x}=\partial_x
S$~\cite{Bohm-1,Bohm-2} or the Langevin equation $\rm{d} \mit{x}=
(\nu \partial_x R /R + \partial_x S) \rm{d} \mit{t} + \nu^{\rm{1/2}}
\rm{d} \eta $~\cite{Nelson}, where $x$ in their case is the
particle's displacement, $\nu$ is proportional to $\hbar$ and $\eta$
is Gaussian white noise. Also we use ensemble of particles to construct the wave
packet, and we combine both stochastic (our initial conditions are
drawn from the Boltzmann-Gibbs distribution) and deterministic
(solution of Newtonian equations) tools. But here the resemblance
ends. In our approach we used the Hilbert transform which is the
most natural way to extend a signal to the complex plane and then we
construct the classical correlation functions which yield the real
and imaginary parts of the wave function. Furthermore, in our
examples, the transformation is used on the individual trajectory
level and only for momentum, so our approach is based on extending
the classical trajectory to the complex plane. Therefore, the wave
function in our case has a classical interpretation in terms of
classical observables without invoking fictitious forces which have
no classical analogue.

With above strategy, we have found that this proposed wave
function's modulus square corresponds to the classical energy and
heat spreading densities. Such densities have been found to exhibit
non-universal shapes (dependent on the phonon dispersion relation),
showing quantitative agreements with the simulation results of
ballistic heat transport in many integrable systems. We have also
proposed the quasi-particle velocity approach to understand the long
time asymptotic behaviour of these ballistic wave packets. This
similarity to quantum walks together with the picture of the quasi
particle velocity approach provides a new perspective to understand
ballistic heat transport. An extension of these ideas for general
nonlinear, nonintegrable systems will provide a ``hydrodynamic
foundation'' of these models~\cite{RW-3,Cai}.

\begin{acknowledgments}
D.X. was supported by the National Natural Science Foundation of
China (Grant No. 11575046); the Natural Science Foundation of Fujian
province, China (Grant No. 2017J06002); the Training Plan Fund for
Distinguished Young researchers from Department of education, Fujian
Province, China; the Qishan Scholar Research Fund of Fuzhou
University, China. E.B. and F.T. were supported by the Israel
Science Foundation.
\end{acknowledgments}

\begin{appendix}
\section{Density profile for the tight-binding Hamiltonian with NN jumps}
\begin{figure}
\begin{centering}
\vspace{-.6cm} \includegraphics[width=8cm]{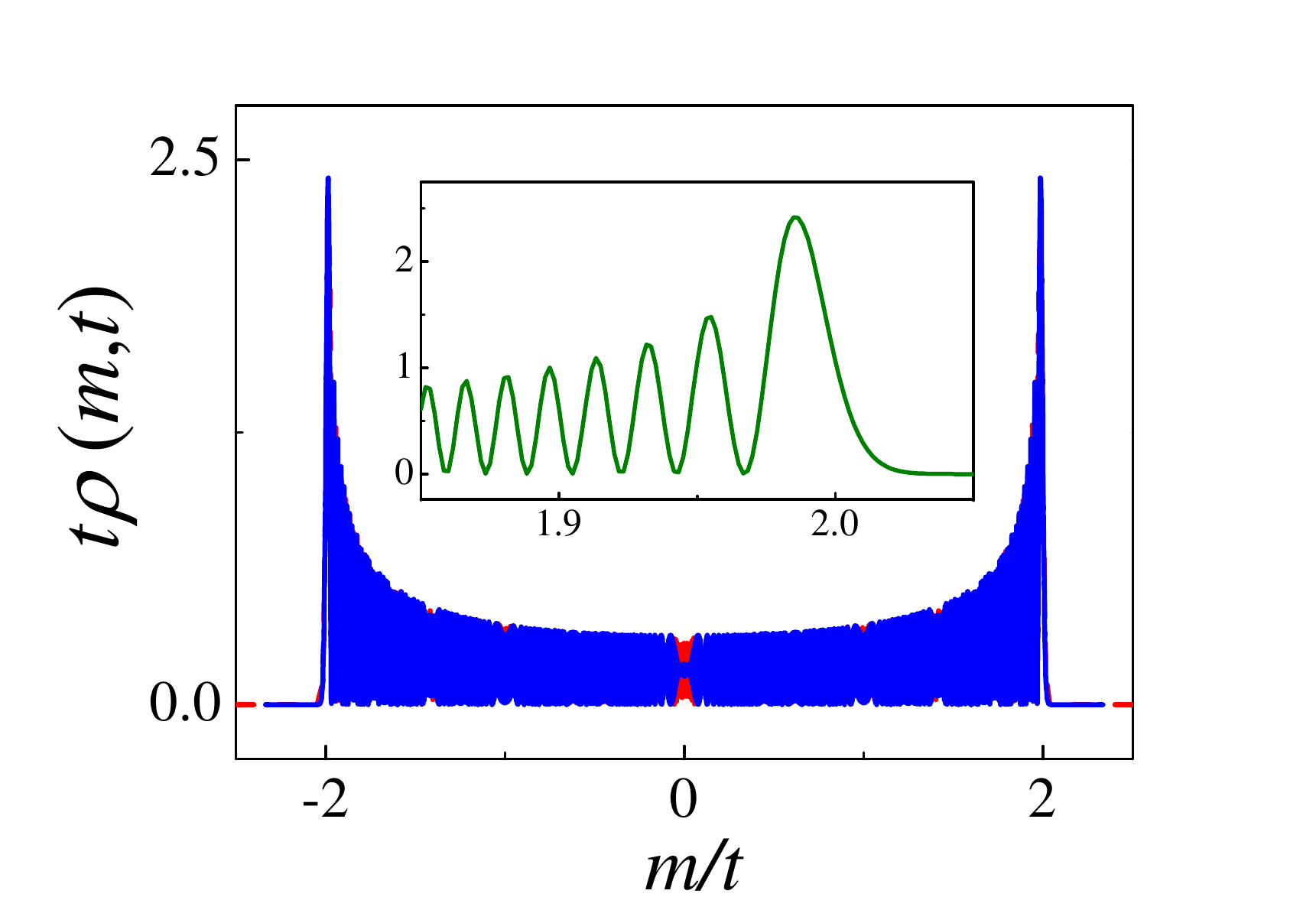} \vspace{-.3cm}
\caption{\label{SFignew1} The rescaled density of the NN
tight-binding quantum walk. Dotted, dashed and solid lines
correspond to $t=200$, $400$ and $600$, respectively. The inset
($t=600$) shows the oscillations (interference) in
detail.}\vspace{-.3cm}
\end{centering}
\end{figure}

As mentioned, the density of the nearest-neighbor (NN) tight-binding
quantum system has three key properties: ballistic scaling, U-shape
and oscillations. All of these are nicely demonstrated in
Fig.~\ref{SFignew1}, where the rescaled density $t\rho(m,t)$ is
plotted against $m/t$. The oscillations are better visible in the
inset.
\section{Simulation detail}
We mainly focus on the following four correlation functions. (i) The
momentum correlation function
\begin{equation} \label{momentum}
\rho_{p}(m,t)=\frac{\langle p_{j}(t) p_{i}(0) \rangle}{\langle
p_{i}(0) p_{i}(0) \rangle},
\end{equation}
and (ii) the cross-correlation function between momentum and its
Hilbert transform
\begin{equation} \label{Pmomentum}
\rho_{\widetilde{p}}(m,t)=\frac{\langle \widetilde{p}_{j}(t)
p_{i}(0) \rangle}{\langle p_{i}(0) p_{i}(0) \rangle}.
\end{equation}
In relation to heat transport, we use (iii) the correlation function
for energy
fluctuations~\cite{Forster,Liquid,Zhao2006,Chen2013,Xiong2016-1,Xiong2016-2}
\begin{equation} \label{EE}
\rho_{E} (m,t)=\frac{\langle \Delta E_{j}(t) \Delta E_{i}(0)
\rangle}{\langle \Delta E_{i}(0) \Delta E_{i}(0) \rangle},
\end{equation}
(iv) and the correlation function for heat energy fluctuations
\begin{equation} \label{QQ}
\rho_{Q} (m,t)=\frac{\langle \Delta Q_{j}(t) \Delta Q_{i}(0)
\rangle}{\langle \Delta Q_{i}(0) \Delta Q_{i}(0) \rangle}.
\end{equation}
Here $m=j-i$; $\langle \cdot \rangle$ represents the spatio-temporal
average; $\Delta \chi \equiv \chi- \langle \chi\rangle$ is the
corresponding quantity's fluctuations. For (i)-(iii) the labels $i$
and $j$ correspond to the labels of particles. The energy $E_i$ is
defined by the sum of kinetic energy $p_i^2/2$ and potential energy
$V$ (which may also depend on the position of other particles).

The heat density $Q_i$ is defined for a finite volume (bin). Its
expression can be derived from basic thermodynamics. For details we
refer to textbooks \cite{Forster,Liquid} and to other publications
\cite{Zhao2006,Chen2013,Xiong2016-1,Xiong2016-2}. The indices for
the heat fluctuation density correspond to bin labels rather than
particle labels. In each bin, we calculate the number of particles
in the bin $M_i$, the energy in the bin $E_i$ and the pressure $F_i$
exerted on the bin. Finally the heat is obtained from $Q_i(t)\equiv
E_i(t)-\frac{(\langle E \rangle +\langle F \rangle)M_i(t)}{\langle M
\rangle}$. Since the system is one dimensional the pressure is equal
to the force and can be calculated from the gradient of the
potential.

For the Toda chain, its general potential is
$V(\xi)=\frac{a}{b}e^{-b \xi}+a \xi + c$, from which one can derive
a modulus dependent dispersion relation~\cite{Toda1979}. For
simulations at finite temperature, we employ the simple form
$V(\xi)=e^{-\xi}+ \xi -1$, i.e., we simply set $a=b=-c=1$. The
modulus $\kappa$ shown in the Toda chain's dispersion relation
$\omega_{q}= \frac{\pi}{K_{1}} / \sqrt{\frac{1}{{\rm{Sn}}^2(K_{1} q
/ \pi)} -1 + \frac{K_{2}}{K_{1}}}$~\cite{Toda1979} is some constant
determined by the parameters $a$, $b$ and $T$ describing the
non-linearity.

For all the simulations, we set both the equilibrium distance
between the particles as well as the lattice constant to unity. So
the number of particles $N$ is equal to the system size. All systems
except the Toda chain have symmetric potentials. Therefore the
average pressure $\langle F \rangle$ calculated from simulations is
always zero. For the Toda chain with an asymmetric potential, the
average pressure depends on temperature. Our simulations indicate
$\langle F \rangle \approx 0.05$, $\langle F \rangle \approx 0.48$
and $\langle F \rangle \approx 1.34$ for $T=0.05$, $T=0.5$ and
$T=1.5$, respectively.

We consider a chain of size $N=4001$ with periodic boundary
conditions. To allow heat and energy fluctuations to actually spread
out, we compute the correlation function up to a lag time of
$t=600$. The heat correlation function was calculated from
discretised chain with $2000$ bins.

We use the stochastic Langevin heat baths~\cite{Report-1,Report-2}
to thermalize the system and to prepare a canonical equilibrium
state with a given temperature. We employ the Runge-Kutta algorithm
of $7$-th to $8$-th order with a time step of $0.05$ to evolve the
system. Each canonical equilibrium system is prepared by evolving
the system for a long enough time ($>10^7$ time units) from properly
assigned random initial states. Then the heat bath is switched off
and we switch to the microcanonical ensemble with fixed energy.
Finally the system is evolved in isolation and we obtain the
correlation information. We used ensembles of circa $8\times10^9$
data points to compute the correlation functions. We also note that
to get the final results of the correlation functions, one should
consider a correction term as suggested in~\cite{Chen2013}. For more
details on the implementation and techniques one can refer
to~\cite{PingHuang}.

To simulate the correlation functions of $\rho_{p}(m,t)$ and
$\rho_{\widetilde{p}}(m,t)$, we first record the time series of
$p_m(t)$. Then $\widetilde{p}_m(t)$ is obtained by using the
numerical Hilbert transform~\cite{Hilbert}. The pairs are then used
to calculate the correlation functions. As an example,
Fig.~\ref{SFignew} plots the results of $p_m(t)$ and
$\widetilde{p}_m(t)$ for the harmonic chain; Fig.~\ref{SFignew2}
depicts the correlation functions of $\rho_{p}(m,t)$ and
$\rho_{\widetilde{p}}(m,t)$ from simulations and compares them with
the wave function's real and imaginary parts.
\begin{figure}
\begin{centering}
\includegraphics[width=8cm]{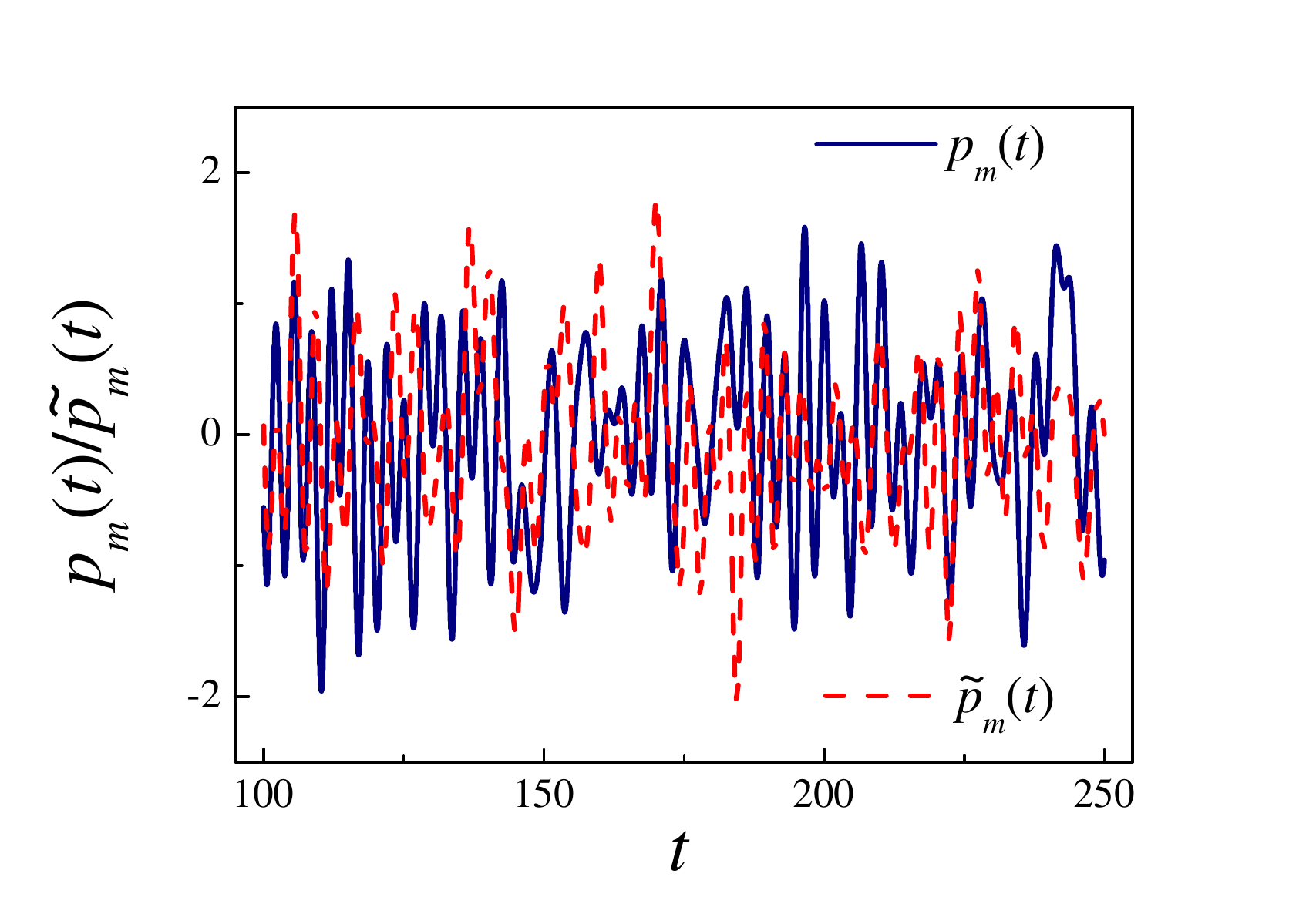}
\vspace{-.6cm} \caption{\label{SFignew} $\widetilde{p}_{m}(t)$ and
$p_m(t)$ versus $t$ for a harmonic chain, here we use the particle
with the label of $m=1000$ for example.}\vspace{-.3cm}
\end{centering}
\end{figure}
\begin{figure}
\begin{centering} \includegraphics[width=8.8cm]{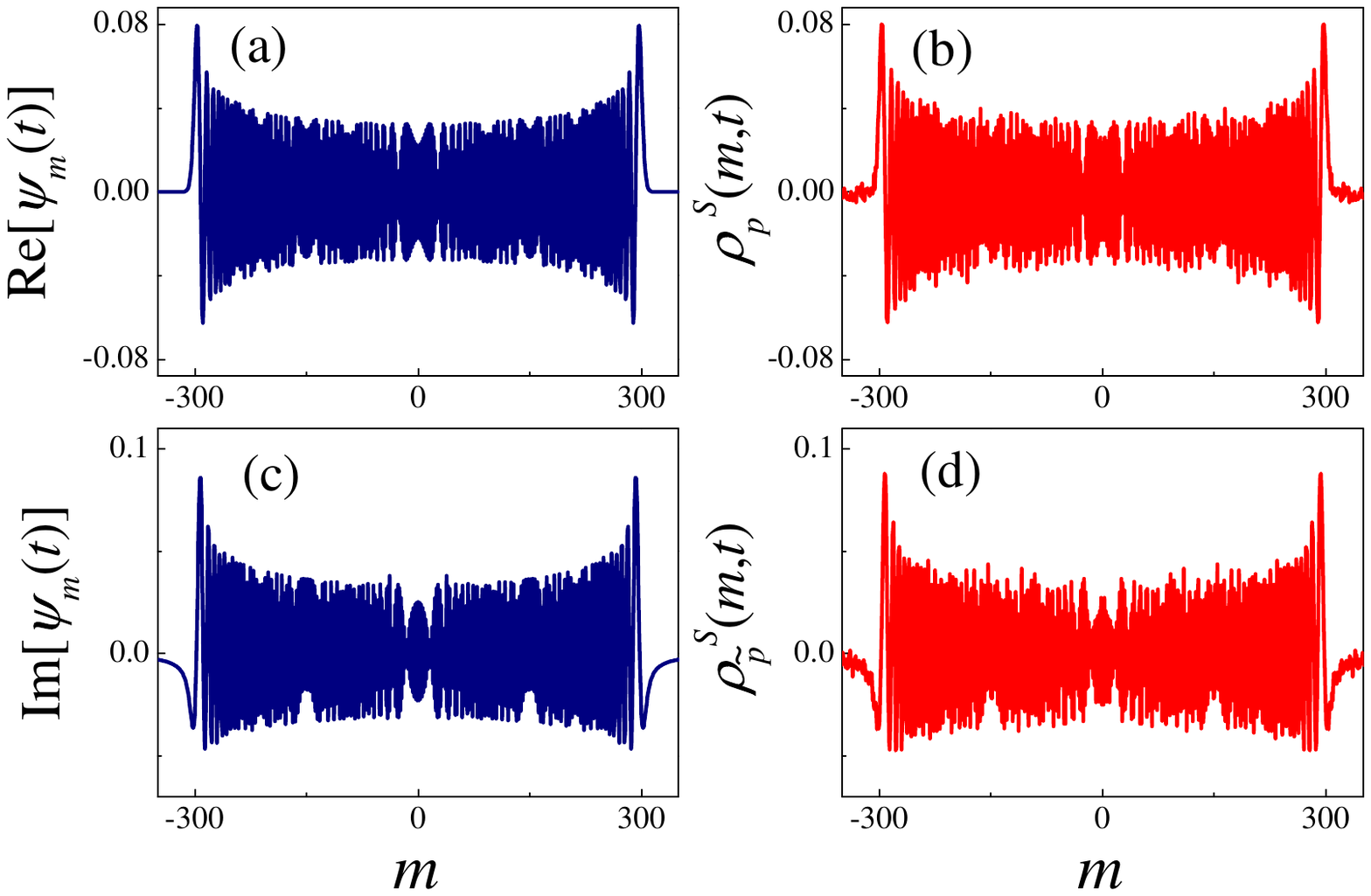}
\vspace{-.8cm} \caption{\label{SFignew2} Comparison of the wave
function's real and imaginary part with $\rho_{p}(m,t)$ and
$\rho_{\widetilde{p}}(m,t)$ from simulations for a harmonic chain
($t=300$ for example).} \vspace{-.3cm}
\end{centering}
\end{figure}
\section{The physical meaning of the wave function}
\subsection{The square of wave function's real part is the kinetic energy correlation function}
\begin{figure}
\begin{centering}
\vspace{-.6cm}\includegraphics[width=8cm]{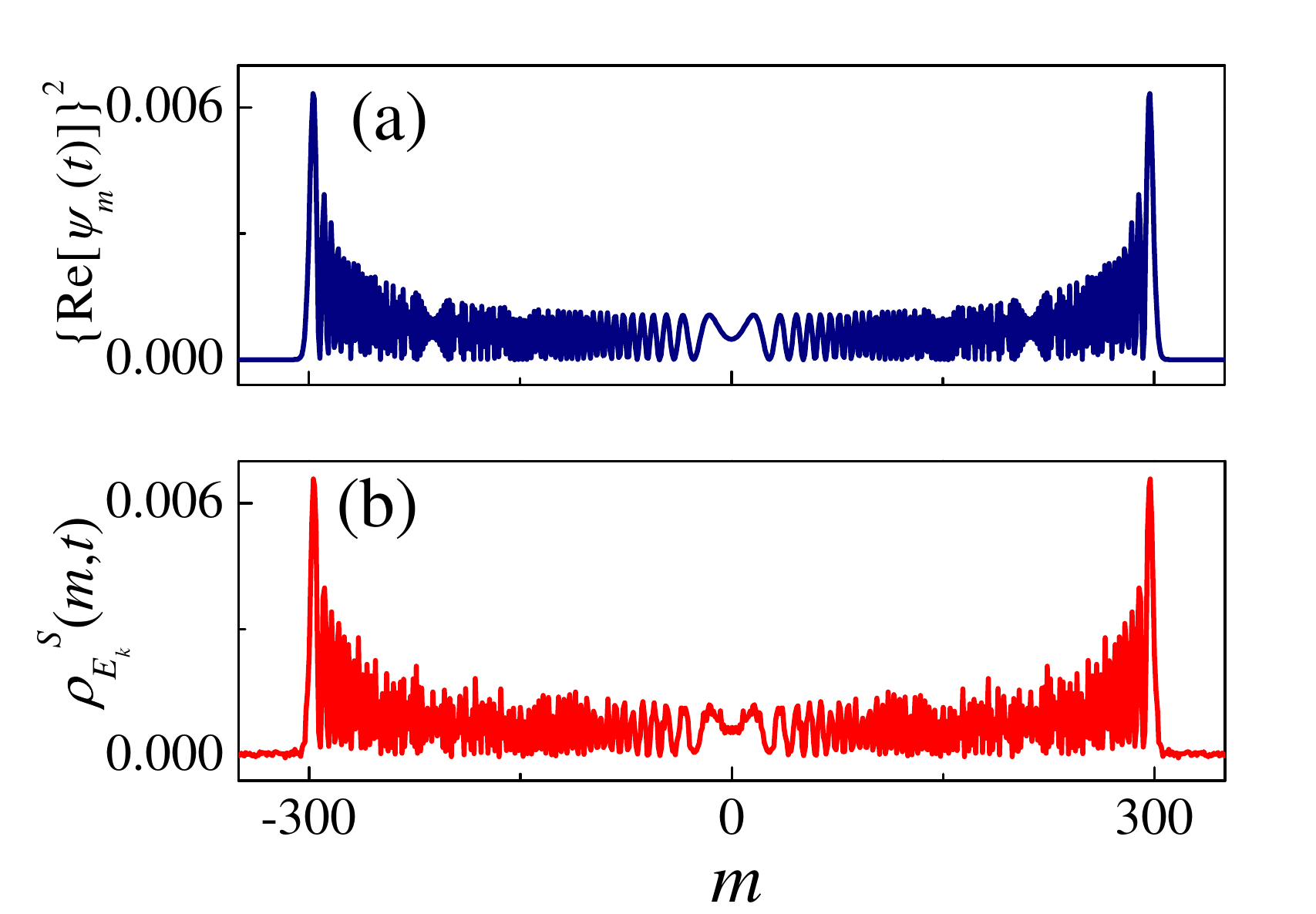} \vspace{-.4cm}
\caption{\label{SFig2} The kinetic energy correlation function for
the harmonic chain: (a) Prediction
$\left\{\rm{Re}[\psi_{\mit{m}}(\mit{t})]\right\}^2$; (b) Simulation
$\rho_{E_k}^{S}(m,t)$, here the results of $t=300$ are plotted.}
\vspace{-.3cm}
\end{centering}
\end{figure}

The kinetic energy of the $m$th particle is $E_m^k = p_m^2/2$,
whence its correlation function $\rho_{E_{k}}(m,t)$ is defined by
\begin{equation}
\rho_{E_{k}} (m,t) = \frac{ \langle \Delta E^{k}_{m}(t) \Delta
E^{k}_{0}(0) \rangle}{\langle \Delta E^{k}_{0}(0) \Delta
E^{k}_{0}(0) \rangle} \nonumber
\end{equation}
\begin{equation}
= \frac{\left \langle \left[ |p_m(t)|^2-\langle |p_m(0)|^2 \rangle
\right] \left[ |p_0(0)|^2-\langle |p_0(0)|^2 \rangle \right] \right
\rangle} {\left \langle \left[ |p_0(0)|^2-\langle |p_0(0)|^2 \rangle
\right] \left[ |p_0(0)|^2-\langle |p_0(0)|^2 \rangle \right] \right
\rangle}.
\end{equation}
Since the system is initially in contact with heat baths, the
initial momenta and the initial positions are jointly normal
distributed. The equipartition conditions stated in the main text
give the variance/covariance of these Gaussian random variables:
$\langle P_k(0)P^*_{k'}(0)\rangle = \delta_{k,k'} k_B T$, $\langle
R_k(0)R^*_{k'}(0)\rangle = \delta_{k,k'} k_B T/
\widetilde{\omega}_k^2$, and $\langle P_k(0)R^*_{k'}(0)\rangle = 0$.
In particular Wick's theorem applies and the fourth moments can be
replaced with the second moments, e.g. $\langle P_k^4(0)\rangle = 3
\langle P_k^2(0) \rangle^2 = 3 (k_B T)^2$. This allows us to compute
the denominator, which is the normalization condition. First note
that it can be simplified to $\langle p_0^4(0) \rangle - \langle
p_0^2(0)\rangle^2 = 3 \langle p_0^2(0) \rangle^2 - \langle
p_0^2(0)\rangle^2 = 2 k_B T$. Hence we have (by also using
translational invariance in time and space):
\begin{equation} \label{EK}
\rho_{E_{k}} (m,t)= \frac{\langle |p_m(t)|^2 |p_0(0)|^2 \rangle
-\langle |p_0(0)|^2 \rangle^2} {2 (k_B T)^2},
\end{equation}
Now we need to find the numerator. We use
Eqs.~(\ref{Pnormal}-\ref{Transform-1}), and employ Einstein's
notation: the summation is carried out over all $k$'s and $l$'s:
\begin{align} \label{Stemp3}
     &
        \langle |p_m(t)|^2 |p_0(0)|^2 \rangle \nonumber
    \\= & \nonumber
        \left\langle
            C_{m,k} \left[
                P_{k}(0) \cos(\widetilde{\omega}_k t)
                - \widetilde{\omega}_k R_{k}(0) \sin (\widetilde{\omega}_k t)
            \right] \times
            \right.
    \\ & \left. \nonumber
            C_{m,k'}^* \left[
                P_{k'}(0) \cos(\widetilde{\omega}_k t)
                - \widetilde{\omega}_k R_{k'}(0) \sin (\widetilde{\omega}_{k'} t)
            \right]
        \right.\times
    \\ & \left.
            C_{0,l} P_{l}(0) C^*_{0,l'} P_{l'}(0)
        \right\rangle.
\end{align}
Using Eq.~\eqref{Transform}, we can substitute $C_{0,l} C_{0,l'}^*$
with $1/N$. Furthermore, since $R_k(0)$ and $P_l(0)$ are
uncorrelated, a lot of the occurring terms actually vanish. We are
left with two big sums:
\begin{align}
    &
        \langle |p_m(t)|^2 |p_0(0)|^2 \rangle \nonumber
    \\ = & \nonumber
        \frac{C_{m,k} C_{m,k'}^*}{N}
        \cos(\widetilde{\omega}_k t)
        \cos(\widetilde{\omega}_{k'} t)
        \left\langle P_k(0)P_{k'}(0)P_l(0)P_{l'}(0) \right\rangle
    \\& + \nonumber
        \frac{C_{m,k} C_{m,k'}^*}{N}
        \widetilde{\omega}_k\sin(\widetilde{\omega}_k t)
        \widetilde{\omega}_{k'}\sin(\widetilde{\omega}_{k'} t)
     \\ & \times
        \left\langle R_k(0)R_{k'}(0)P_l(0)P_{l'}(0) \right\rangle
\end{align}
We apply Wick's theorem on the remaining expectations: $\left\langle
P_k(0)P_{k'}(0)P_l(0)P_{l'}(0) \right\rangle = (k_BT)^2 (
\delta_{l,l'}\delta_{k,k'} +  \delta_{k,l'}\delta_{k',l} +
\delta_{k,l}\delta_{k',l'})$. The other expectation is easier,
because $R_k(0)$ and $P_l(0)$ are not correlated: $\left\langle
R_k(0)R_{k'}(0)P_l(0)P_{l'}(0) \right\rangle = \delta_{k,k'}
\delta_{l,l'} (k_BT)^2 / \widetilde{\omega}_k^2$. We obtain:
\begin{align}
    & \nonumber
        \langle |p_m(t)|^2 |p_0(0)|^2 \rangle
    \\ = & \nonumber
        \frac{C_{m,k} C_{m,k'}^*}{N}
        \cos(\widetilde{\omega}_k t)
        \cos(\widetilde{\omega}_{k'} t) \times
    \\ & \nonumber
        (k_BT)^2 ( \delta_{l,l'}\delta_{k,k'} +  \delta_{k,l'}\delta_{k',l} + \delta_{k,l}\delta_{k',l'})
        +
    \\ & \nonumber +
        \frac{C_{m,k} C_{m,k'}^*}{N}
        \widetilde{\omega}_k\sin(\widetilde{\omega}_k t)
        \widetilde{\omega}_{k'}\sin(\widetilde{\omega}_{k'} t)
        \delta_{k,k'} \delta_{l,l'} \frac{(k_BT)^2}{\widetilde{\omega}_k^2}
    \\ = &
        (k_BT)^2 \left[
            1
            +
            \frac{2}{N} C_{m,k} C_{m,k'}^*
            \cos(\widetilde{\omega}_k t)
            \cos(\widetilde{\omega}_{k'} t)
        \right]
\end{align}
The sum over $l$ and $l'$ cancels some of the $1/N$ factors. The
terms with $\delta_{k,k'}$ lead to $\cos^2(\widetilde{\omega}_k t) +
\sin^2(\widetilde{\omega}_k t) = 1$, which bears the first summand
of the result. We obtain the kinetic energy correlation function
from plugging the last equation into Eq.~\eqref{EK}:
\begin{align} \label{Stemp5}\
        \rho_{E_{k}} (m,t)
    = & \nonumber
        \frac{1}{N} \left|\sum_{k=0}^{L} C_{m,k} \cos(\widetilde{\omega}_k t)\right|^2
    \\ = &
        \left| \frac{1}{N} \sum_{k=0}^{L} \exp (\frac{\rm{i} 2 \pi\mit{m} k} {N}) \cos (\widetilde{\omega}_{k} t) \right |^2,
\end{align}
where we used $C_{m,k}=\frac{1}{\sqrt{N}} \exp \left(2 \pi
\rm{i}\frac{\mit{m} k} {\mit{N}} \right)$. Finally, taking $N = L+1
\rightarrow \infty$, we find that \eqref{Stemp5} is actually
\begin{equation}
\rho_{E_{k}} (m,t)=
\left\{\rm{Re}[\psi_{\mit{m}}(\mit{t})]\right\}^2=\left[ \frac{1}{2
\pi} \int_{- \pi}^{\pi} \cos (q m) \cos (\omega_q t) \rm{d}
\mit{q}\right] ^{\rm{2}},
\end{equation}
the square of the wave function's real part
$\left\{\rm{Re}[\psi_{\mit{m}}(\mit{t})]\right\}^{\rm{2}}$. To
verify our proof, we compared the real part with the kinetic
energy's correlation function in Fig.~\ref{SFig2} for the harmonic
chain.

An alternative way to prove the equivalence between $\left\{\rm{Re}[\psi_{\mit{m}}(\mit{t})]\right\}^{\rm{2}}$ and $\rho_{E_{k}} (m,t)$ relies only on the Gaussian property of the momenta. Considering two Gaussian random variables
$X$, $Y$ with zero mean ($\langle X \rangle =0$; $\langle Y \rangle
=0$) and variance ($\langle X^2 \rangle=\sigma_X^2$; $\langle Y^2
\rangle=\sigma_Y^2$). We write their joint
distribution as
\begin{align}
    & \nonumber
    P(X,Y)=\frac{1} {2 \pi \sigma_X \sigma_Y \sqrt{1-\varrho^2}} \times \\
    & \nonumber
     \exp \left[ -\frac{1} {2(1- \varrho^2)} \left(
\frac{X^2}{\sigma_X^2} + \frac{Y^2}{\sigma_Y^2} -\frac{2 \varrho X
Y} {\sigma_X \sigma_Y} \right) \right],
\end{align}
where $\varrho=\frac{\langle XY \rangle} {\sigma_X \sigma_Y}$ is
their correlation coefficient. From this joint distribution, we have
\begin{eqnarray}
\langle X^2 Y^2 \rangle &=& \langle X^2  \rangle \langle Y^2 \rangle +2 \langle X Y \rangle^2 \nonumber \\
&=& \langle X^2  \rangle \langle Y^2 \rangle +2 \sigma_X^2 \sigma_Y^2 \varrho^2 \nonumber \\
&=& \langle X^2  \rangle \langle Y^2 \rangle (1+2 \varrho^2),
\end{eqnarray}
which then helps us to calculate $\langle |p_m(t)|^2 |p_0(0)|^2
\rangle$:
\begin{eqnarray}
\langle |p_m(t)|^2 |p_0(0)|^2 \rangle &=& \langle |p_m(t)|^2  \rangle \langle |p_0(0)|^2 \rangle (1+2 \varrho^2) \nonumber \\
&=& (k_B T)^2 (1+2 \varrho^2).
\end{eqnarray}
Obviously, $\langle |p_m(t)|^2 \rangle=\langle |p_0(0)|^2 \rangle=k_B T$, and $\varrho= \frac {\langle p_m(t) p_0(0) \rangle} {k_B
T}=\rho_p (m,t)$ is just the momentum correlation. So, we obtain
\begin{eqnarray} \label{KineticE}
\rho_{E_{k}} (m,t) &=& \frac{(k_B T)^2 \left\{1+2 [\rho_p
(m,t)]^2\right\} -(k_B T)^2} {2 (k_B T)^2} \nonumber \\
&=& [\rho_p
(m,t)]^2=\left\{\rm{Re}[\psi_{\mit{m}}(\mit{t})]\right\}^2.
\end{eqnarray}
\subsection{Stretch correlation function}
Define the NN stretch as $s_m=\Delta r_m=r_{m+1}-r_{m}$, the stretch
correlation function is
\begin{equation}
\rho_{s} (m,t)=\frac{\frac{1}{2} \langle s_m(t) s_0^{*}(0)+  s_0(t)
s_m^{*}(0) \rangle} {\langle |s_0(0)|^2 \rangle}.
\end{equation}
Repeating the steps of the last section, we get:
\begin{equation}
\rho_{s} (m,t)=\frac{C_{s} (m,t)}{\langle |s_{0}(0) |^2 \rangle}=
\frac{ \int_{-\pi}^{\pi} \rm{cos} (\mit{\omega_q t}) \rm{cos}
(\mit{q m}) \frac{\rm{1}- \rm{cos} (\mit{q})} {\omega_q^{\rm{2}}}
\rm{d} \mit{q}} {\int_{-\pi}^{\pi} \frac{\rm{1}- \rm{cos} (\mit{q})}
{\omega_q^{\rm{2}}} \rm{d} \mit{q}}
\end{equation}
with
\begin{equation} \label{Stemp7}
\langle |s_{0}(0) |^2 \rangle= \frac{k_B T}{2 \pi} \int_{-\pi}^{\pi}
\frac{\rm{2}-\rm{2} \cos (\mit{q})} {\omega_q^{\rm{2}}} \rm{d}
\mit{q}.
\end{equation}
The non-normalized numerator is
\begin{align}
    & \nonumber
        C_{s} (m,t)
    \\ = & \nonumber
        \frac{1}{2} \langle s_m(t) s_0^{*}(0)+ s_0(t) s_m^{*}(0) \rangle
    \\ = & \nonumber
        \frac{k_B T}{2 \pi} \int_{-\pi}^{\pi} \frac{ \rm{cos} (\mit{\omega_q t})} {\omega_q^{\rm{2}}}
    \\ & \times \nonumber
        \left \{2 \rm{cos} (\mit{q m}) -\rm{cos} (\mit{q} m+ q ) - \rm{cos} (\mit{q} m- q) \right\} \rm{d} \mit{q}
    \\ = &
        \frac{k_B T}{2 \pi} \int_{-\pi}^{\pi} \rm{cos} (\mit{\omega_q t}) \rm{cos} (\mit{q m}) \frac{\rm{2}-2 \rm{cos} (\mit{q})}{\omega_q^{\rm{2}}} \rm{d} \mit{q}.
\end{align}
Now it is interesting to find that $C_{s} (m,t)$ is related to the
wave function via
\begin{equation}
\frac{\rm{d}^2}{\rm{d}\mit{t}^2} \left[ \frac{C_{s } (m,t)} {k_B
T}\right]=\rm{Re} \left \{ \psi_{\mit{m}+ \rm{1}} (\mit{t}) +
\psi_{\mit{m}- \rm{1} } (\mit{t}) - \rm{2} \psi_{\mit{m}} (\mit{t})
\right \}.
\end{equation}

\begin{figure}
\begin{centering}
\vspace{-.6cm} \includegraphics[width=8.8cm]{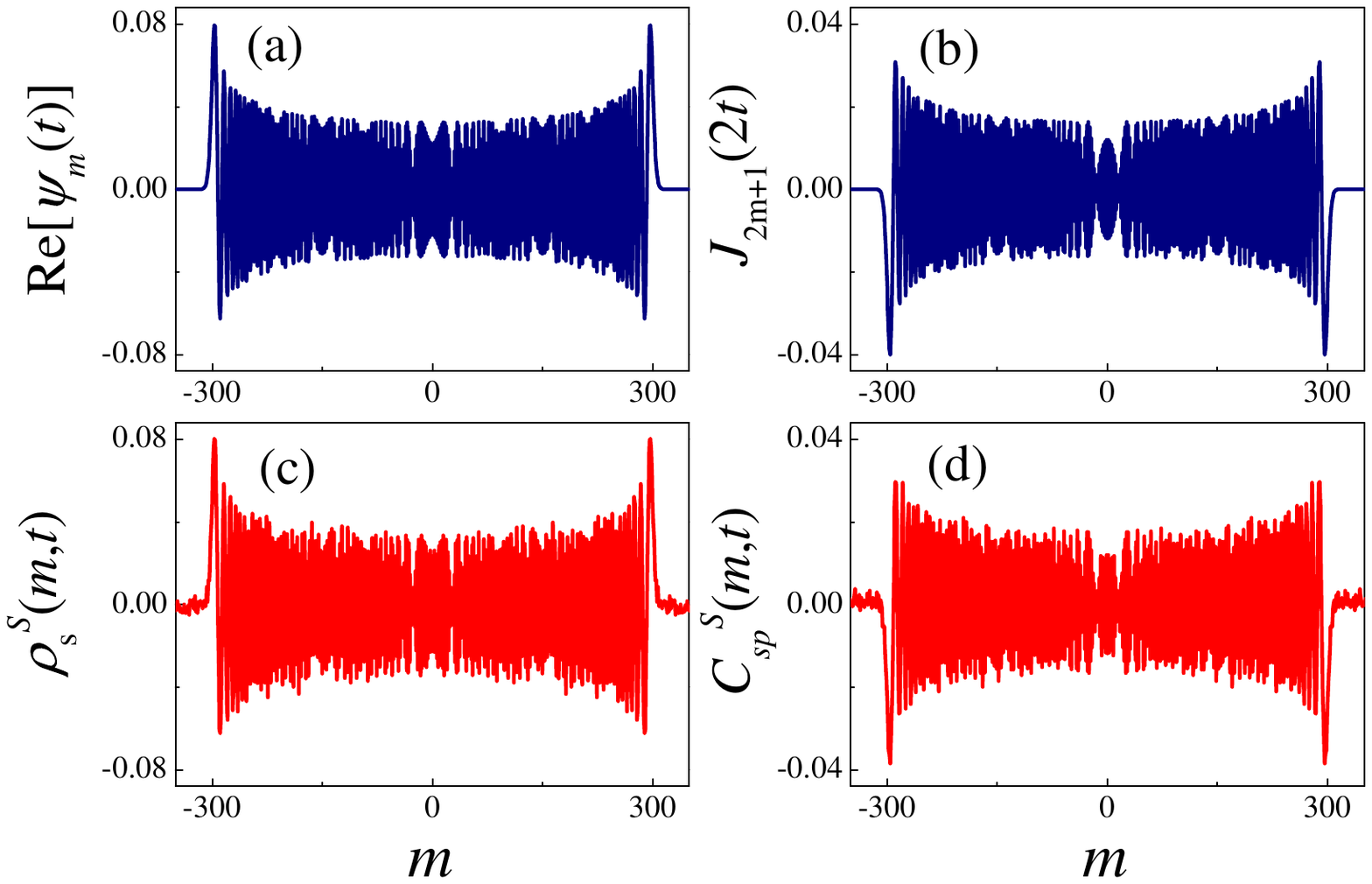} \vspace{-.8cm}
\caption{\label{SFig3} The stretch correlation function [(a) and
(c)], the stretch-momentum cross-correlation function [(b) and (d)]
for a harmonic chain: (a)-(b) from predictions; (c)-(d) from
simulations ($t=300$).} \vspace{-.3cm}
\end{centering}
\end{figure}

Take the harmonic chain as an example. Its dispersion relation is
$\omega_q = 2 \left|\sin \left(\frac{q}{2}\right)\right| = \sqrt{2-2
\rm{cos} (\mit{q})}$. Inserting this expression one gets
\begin{eqnarray} \label{stretchC}
\rho_{s} (m,t)&=&\frac{\rm{1}}{\rm{2} \mit{\pi}} \int_{-\pi}^{\pi}
\cos (q m) \cos (\omega_q t)
\rm{d} \mit{q} \nonumber \\
&=& J_{2m} (2t)= \rm{Re} [\mit{\psi}_{m}(t)],
\end{eqnarray}
which is verified by simulations in Fig.~\ref{SFig3}(a) and (c).
\subsection{Stretch-momentum cross-correlation function}
The stretch-momentum cross-correlation function is defined as
\begin{equation}
C_{s p} (m,t)=\frac{1}{2} \langle s_m(t) p_0^{*}(0)+ s_m^{*}(t)
p_0(0) \rangle.
\end{equation}
Following the similar steps of the above, one finds
\begin{align}
    & \nonumber
        C_{s p} (m,t)
    \\ = &
       \frac{k_B T}{2 \pi} \int_{-\pi}^{\pi} \frac{ \rm{sin} (\mit{\omega_q t})} {\omega_q} \left \{\rm{cos} (\mit{q} m+ q ) - \rm{cos} (\mit{q m}) \right\} \rm{d} \mit{q},
\end{align}
which is related to the wave function by
\begin{equation}
\frac{\rm{d}}{\rm{d}\mit{t}} \left[ \frac{C_{s p} (m,t)} {k_B
T}\right]=\rm{Re} \left [ \psi_{\mit{m}+ \rm{1}}
(\mit{t})-\psi_{\mit{m}} (\mit{t}) \right].
\end{equation}
Inserting the harmonic chain's dispersion relation $\omega_q = 2
\left|\sin \left(\frac{q}{2}\right)\right| = \sqrt{2-2 \rm{cos}
(\mit{q})}$, we obtain
\begin{equation}
C_{s p} (m,t)=J_{2m+1}(2t).
\end{equation}
Similarly, it is easy to find that the momentum-stretch cross-correlation function is
\begin{equation}
C_{p s} (m,t)=J_{2m-1}(2t).
\end{equation}
In Fig.~\ref{SFig3}(b) and (d), we compare the prediction of $C_{s p} (m,t)$ with
simulation. As expected, they agree well with each other.
\subsection{Potential energy and total energy correlation functions}
Following reference~\cite{Dharnew}, one
can define the potential energy for harmonic chain as
\begin{equation}
E_m^{p}=\frac{{(r_{m+1}-r_m)}^2}{2}=\frac{(\Delta
r_m)^2}{2}=\frac{s_m^2}{2}.
\end{equation}
Under this definition, the potential energy correlation function is
\begin{eqnarray} \label{PotentialD}
\rho_{E_{p}} (m,t) &=&\frac{ \langle \Delta E^{p}_{m}(t) \Delta
E^{p}_{0}(0) \rangle}{\langle \Delta E^{p}_{0}(0) \Delta
E^{p}_{0}(0) \rangle} \nonumber \\
&=&\frac{\left \langle \left[ E^{p}_{m}(t)-\langle E^{p}_{m}(t)
\rangle \right] \left[ E^{p}_{0}(0)-\langle E^{p}_{0}(0) \rangle
\right] \right \rangle} {\left \langle \left[ E^{p}_{0}(0)-\langle
E^{p}_{0}(0) \rangle \right] \left[ E^{p}_{0}(0)-\langle
E^{p}_{0}(0) \rangle \right]
\right \rangle} \nonumber \\
&=&\frac{\langle E^{p}_{m}(t) E^{p}_{0}(0) \rangle - \langle
E^{p}_{0}(0) \rangle^2}{\left \langle [E^{p}_{0}(0)]^2
\right \rangle - \langle E^{p}_{0}(0) \rangle^2} \nonumber \\
&=& \frac{\langle [s_m(t)]^2 [s_{0}(0)]^2 \rangle - \langle
[s_{0}(0)]^2 \rangle^2}{\left
\langle  [s_{0}(0)]^4 \right \rangle - \left \langle [s_{0}(0)]^2 \right \rangle^2} \nonumber \\
&=& \rho_{s^2} (m,t).
\end{eqnarray}
Since $r_m(t)$ and $r_0(0)$ are the Gaussian random variables, so are
$s_m(t)$ and $s_0(0)$. Given $\rho_s(m,t)$, $\rho_{s^2} (m,t)$ can
be derived by using the alternative argument of appendix C1. From above,
we already know $\langle |s_{0}(0) |^2 \rangle= \frac{k_B T}{2
\pi} \int_{-\pi}^{\pi} \frac{\rm{2}-\rm{2} \cos (\mit{q})}
{\omega_q^{\rm{2}}} \rm{d} \mit{q}=k_B T$, for the harmonic chain. Due to the translation
invariance, it can be expected $\langle |s_{m}(t) |^2 \rangle
=\langle |s_{0}(0) |^2 \rangle = k_B T$. So
\begin{equation}
\left \langle [s_{0}(0)]^2 \right \rangle^2=  (k_B T)^2,
\end{equation}
and
\begin{equation}
\left \langle [s_{0}(0)] \right \rangle^4= 3 (k_B T)^2.
\end{equation}
Now similar to appendix C1,
\begin{equation}
\langle [s_m(t)]^2 [s_{0}(0)]^2 \rangle=(k_B T)^2 \{1+2
[\rho_{s^2} (m,t)]^2 \}.
\end{equation}
Accordingly,
\begin{eqnarray}
\rho_{E_{p}} (m,t) &=& \rho_{s^2} (m,t) \nonumber \\
&=& \frac{(k_B T)^2 \{1+2 [\rho_{s} (m,t)]^2 \}-(k_B T)^2} {3 (k_B T)^2- (k_B T)^2} \nonumber \\
&=& [\rho_{s} (m,t)]^2.
\end{eqnarray}
Finally, in view of Eq.~\eqref{stretchC}, for the harmonic chain
\begin{equation} \label{PotentialE}
\rho_{E_{p}} (m,t)=[J_{2m}(2t)]^2=\{\rm{Re}
[\mit{\psi}_{m}(t)]\}^{\rm{2}},
\end{equation}
which is related to the wave function.

Next, we deal with the total energy correlation function. The total
energy is defined by
\begin{equation}
E_{m}= E^{k}_{m}+ E^{p}_{m}=\frac{p_m^2}{2} + \frac{s_m^2}{2}.
\end{equation}
Under this definition, its non-normalized correlation function is
\begin{align}
    & \nonumber
        C_{E} (m,t)
    \\ =& \nonumber
       \left \langle \Delta E_m(t) \Delta E_0(0) \right \rangle
    \\=& \nonumber
        \left \langle \left [E_m(t) - k_B T \right]  \left[E_0(0)-k_B T \right] \right \rangle
    \\=& \nonumber
        \left \langle \left [E^{k}_{m}(t)+ E^{p}_{m}(t) - k_B T \right]  \left[E^{k}_{0}(0)+E^{p}_{0}(0)-k_B T \right] \right \rangle
    \\=& \nonumber
        \left \langle \left [E^{k}_{m}(t)+ E^{p}_{m}(t) - k_B T/2 -k_B T/2 \right] \times \right.
    \\ & \left. \nonumber
        \left[E^{k}_{0}(0)+E^{p}_{0}(0)- k_B T/2 -k_B T/2 \right] \right \rangle
    \\=& \left \langle \left [ \Delta E^{k}_{m}(t) + \Delta  E^{p}_{m}(t) \right] \left [ \Delta E^{k}_{0}(0) + \Delta  E^{p}_{0}(0) \right] \right \rangle.
\end{align}
The terms of $k_B T$ and $k_B T/2$ are because the equipartition
conditions tell us $\langle E_0(0) \rangle =\langle E_m(t) \rangle
=k_B T$, $\langle E^{k}_0(0) \rangle =\langle E^{k}_m(t) \rangle
=k_B T/2$, and $\langle E^{p}_0(0) \rangle =\langle E^{p}_m(t)
\rangle =k_B T/2$. Now $C_{E} (m,t)$ can be divided into the
following four terms:
\begin{equation}
C_{E} (m,t)=F_1+F_2+F_3+F_4,
\end{equation}
with
\begin{equation}
F_1= \left \langle \Delta E_m^k(t)  \Delta E_0^k(0) \right \rangle,
\end{equation}
\begin{equation}
F_2= \left \langle \Delta E_m^p(t)  \Delta E_0^p(0) \right \rangle,
\end{equation}
\begin{equation}
F_3= \left \langle \Delta E_m^k(t)  \Delta E_0^p(0) \right \rangle,
\end{equation}
and
\begin{equation}
F_4=\left \langle \Delta E_m^p(t)  \Delta E_0^k(0) \right \rangle.
\end{equation}
Due to~\eqref{KineticE},~\eqref{PotentialD}, and~\eqref{PotentialE},
it is easy to find $F_1=F_2=\frac{(k_B T)^2}{2}
[J_{2m}(2t)]^2=\frac{(k_B T)^2}{2} \{\rm{Re}
[\mit{\psi}_{m}(t)]\}^{\rm{2}}$. Applying the similar normal mode
analysis together with the alternative argument of appendix C1, we
can straightforwardly obtain $F_3=\frac{(k_B T)^2}{2}
[J_{2m-1}(2t)]^2=\frac{(k_B T)^2}{2} [C_{ps}(m,t)]^2$ and
$F_3=\frac{(k_B T)^2}{2} [J_{2m+1}(2t)]^2=\frac{(k_B T)^2}{2}
[C_{sp}(m,t)]^2$. Therefore, the non-normalized total energy
correlation function is~\cite{Dharnew}
\begin{align}
    & \nonumber
        C_{E} (m,t)=\frac{(k_B T)^2}{2} \times
    \\ &
        \{\{\rm{Re}
[\mit{\psi}_{m}(t)]\}^{\rm{2}}
+[C_{ps}(m,t)]^{\rm{2}}+[C_{sp}(\mit{m},t)]^{\rm{2}} \},
\end{align}
which then is related to the wave function.
\begin{figure}
\begin{centering}
\vspace{-.6cm} \includegraphics[width=8.8cm]{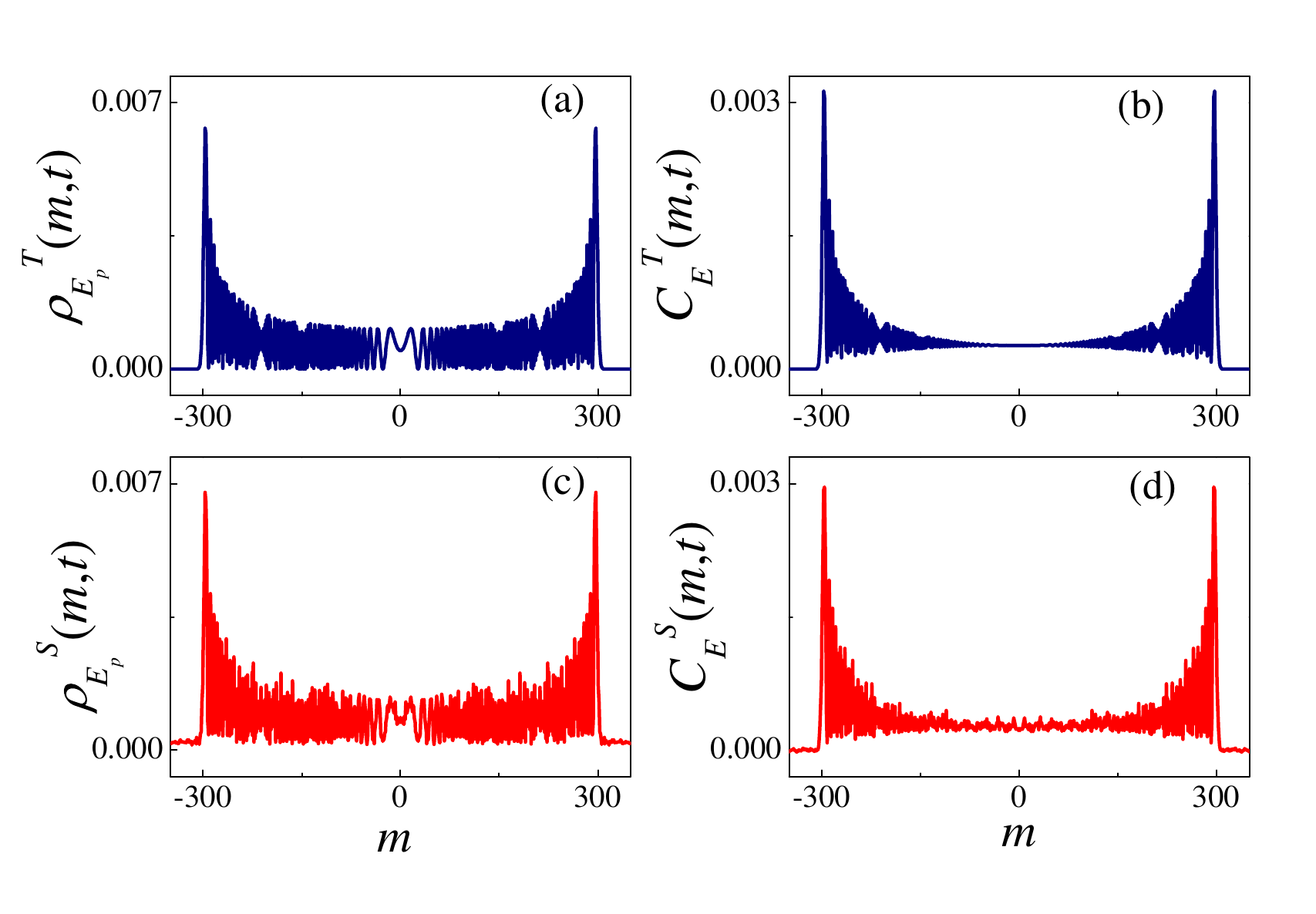} \vspace{-.8cm}
\caption{\label{SFignew} The potential correlation function [(a) and
(c)], the non-normalized total energy correlation function [(b) and
(d)] for a harmonic chain: (a)-(b) from predictions; (c)-(d) from
simulations ($t=300$).} \vspace{-.3cm}
\end{centering}
\end{figure}

Figure~\ref{SFignew} presents both the predictions and simulations
for $\rho_{E_p}(m,t)$ and $C_{E}(m,t)$. As can be seen, they agree
well with each other.
\begin{figure}
\begin{centering}
\vspace{-.6cm} \includegraphics[width=8cm]{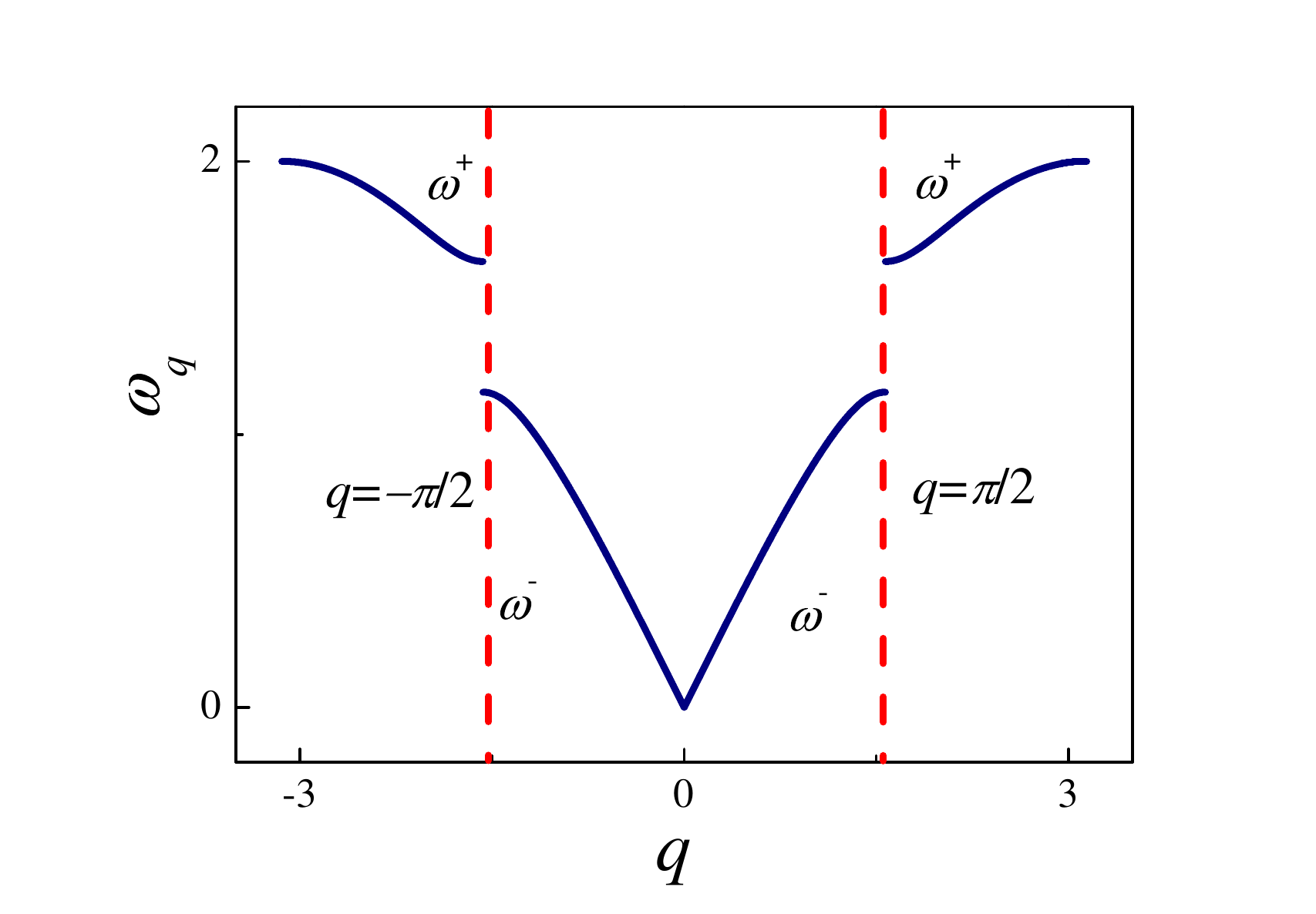} \vspace{-.6cm}
\caption{\label{SFig4} (Colour online) The phonon dispersion
relation for Model II.}\vspace{-.3cm}
\end{centering}
\end{figure}
\section{the system with two branches of phonons}
\begin{figure}
\begin{centering}
\includegraphics[width=8cm]{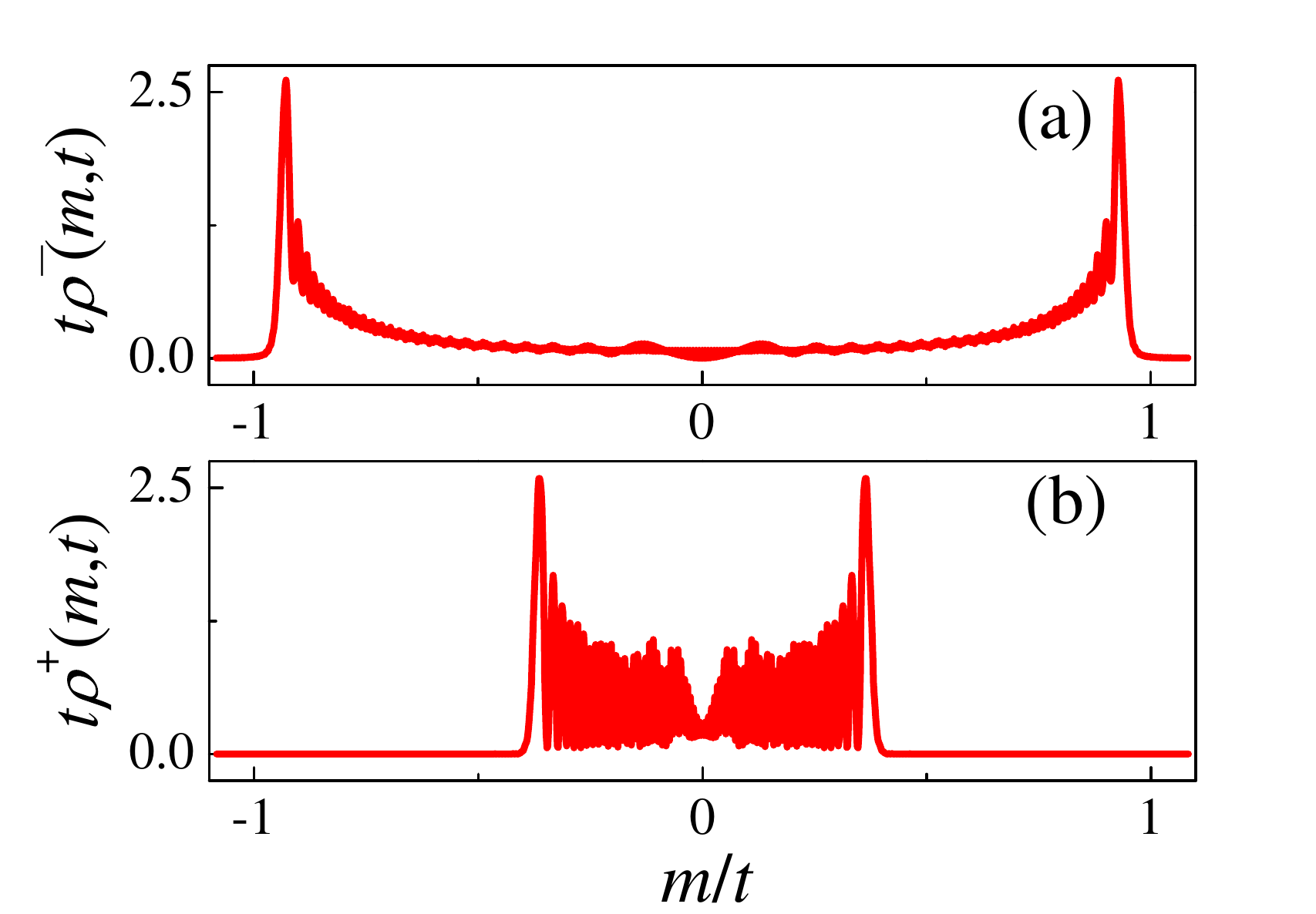} \vspace{-.4cm}
\caption{\label{SFig5}(Colour online) The rescaled $\rho^{-} (m,t)$
(a) and $\rho^{+} (m,t)$ (b) for Model II ($t=600$).} \vspace{-.3cm}
\end{centering}
\end{figure}
We demonstrate here how to derive the density for Model II. First we
plot its phonon dispersion relation in Fig.~\ref{SFig4}. One can see
that this dispersion relation is divided at $q= \pm \frac{\pi}{2}$
into two parts, namely the acoustic and optical phonons. When one
obtains the density $\rho (m,t) = \left|\frac{1}{2 \pi}
\int_{-\pi}^{\pi} e^{\rm{i} \left(\mit{m} \mit{q}- \omega_{q}
\mit{t} \right)} \rm{d} \mit{q}\right|^{2}$, naturally the
integration should be piecewise, hence the density is
\begin{equation}
\rho (m,t)=\left|\psi_{m}^{-}(t)+ \psi_{m}^{+}(t)\right|^{2}
\end{equation}
with
\begin{equation}
\psi_{m}^{-}(t)= \frac{1}{2 \pi}
\int_{-\frac{\pi}{2}}^{\frac{\pi}{2}} e^{\rm{i} \left(\mit{m}
\mit{q}- \omega_{q}^{-} \mit{t} \right)} \rm{d} \mit{q}
\end{equation}
and
\begin{equation}
\psi_{m}^{+}(t)=\frac{1}{2 \pi} \left[
\int_{-\pi}^{-\frac{\pi}{\rm{2}}} e^{\rm{i} \left(\mit{m} \mit{q}-
\omega_{q}^{+} \mit{t} \right)} \rm{d} \mit{q} +
\int_{\frac{\pi}{\rm{2}}}^{\pi}  e^{\rm{i} \left(\mit{m} \mit{q}-
\omega_{q}^{+} \mit{t} \right)} \rm{d} \mit{q} \right].
\end{equation}
\begin{figure}
\begin{centering}
\includegraphics[width=8.8cm]{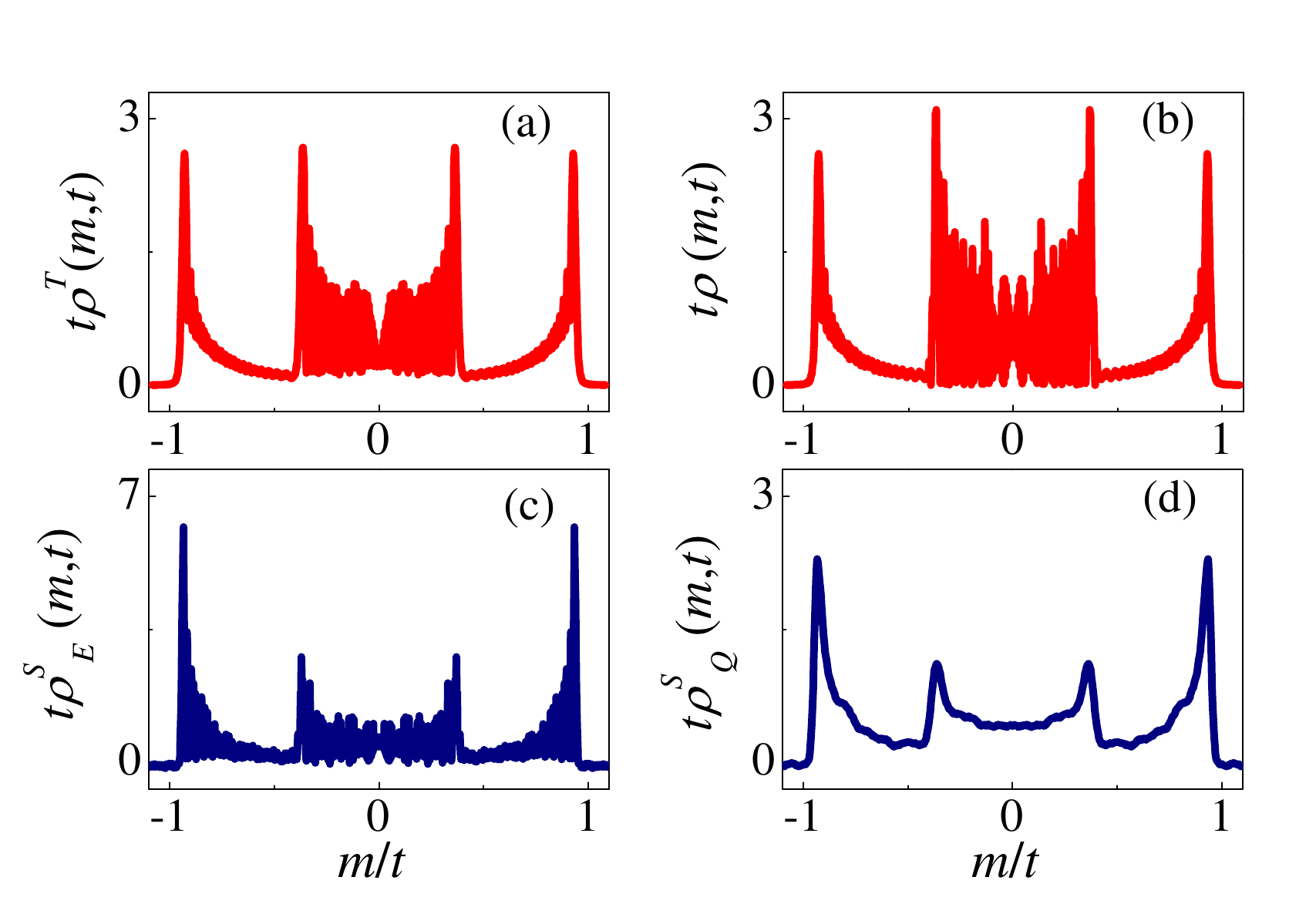} \vspace{-.8cm}
\caption{ (Colour online) The rescaled $\rho^{T} (m,t)$ (a), $\rho
(m,t)$ (b), $\rho^{S}_{E} (m,t)$ (c) and $\rho^{S}_{Q} (m,t)$ (d),
for Model II ($t=600$).\label{Sfig6}}
\end{centering}
\end{figure}

In view of this fact, it would be interesting to take the
contributions of acoustic and optical phonons into account
independently, i.e., let $\rho^{-}
(m,t)=\left|\psi_{m}^{-}(t)\right|^{2}$ [see Fig.~\ref{SFig5}(a)]
and $\rho^{+} (m,t)=\left|\psi_{m}^{+}(t)\right|^{2}$ [see
Fig.~\ref{SFig5}(b)], respectively, then set $\rho^{T}
(m,t)=\rho^{-} (m,t)+\rho^{+} (m,t)$. In Fig.~\ref{Sfig6} we compare
the result of $\rho^{T} (m,t)$ with $\rho (m,t)$ and simulations.
Fortunately, we find that in this particular model, the shape of
$\rho^{T} (m,t)$ [Fig.~\ref{Sfig6}(a)] coincides nicely with $\rho
(m,t)$ [Fig.~\ref{Sfig6}(b)] and also with the simulations
[Figs.~\ref{Sfig6}(c)-(d)]. We therefore argue that one would be
able to separate the contributions of acoustic and optical phonons.
This may stimulate possible applications for the design of phononics
devices~\cite{Note}.
\end{appendix}


\begin{thebibliography}{99}
\bibitem{Note} In certain quantum walks one may use an electron's spin as a flipping device. Possibly the two types of phonons could be used for similar aims. In the Appendix D, we also explain that the independent contributions of acoustic and optical phonons can be identified in practice.
\end{thebibliography}
\end{document}